\documentclass[twocolumn,dvipsnames]{aastex62}
\pdfoutput=1 
\usepackage{amsmath,amstext}
\usepackage{apjfonts}
\usepackage[figure,figure*]{hypcap}
\usepackage{ulem}
\usepackage{xspace}

\usepackage{hyperref}
\hypersetup{
    colorlinks=true,
    linkcolor=blue,
    filecolor=magenta,      
    urlcolor=cyan,
    pdftitle={Overleaf Example},
    pdfpagemode=FullScreen,
    }

\newcommand{\harm}{\xspace{HARM3D}\xspace}

\shorttitle{HARM3D+NUC: GRMHD, nuclear tables and neutrino leakage}
\shortauthors{Murguia-Berthier et al.}

\newcommand{\beq}[1]{\begin{equation} #1 \end{equation}}
\newcommand{\beqa}[1]{\begin{eqnarray} #1 \end{eqnarray}}

\newcommand{\rmin}{r_\mathrm{min}}
\newcommand{\rmax}{r_\mathrm{max}}

\newcommand{\dF}{{^{^*}\!\!F}}

\newcommand{\bF}{{\bf F}}
\newcommand{\bU}{{\bf U}}

\newcommand{\del}{{\partial}}
\newcommand{\bsq}{{{||b||^2}}}
\newcommand{\CTB}{{{\mathcal{B}}}}

\newcommand{\prim}{{{\mathbf{P}}}}

\newcommand{\GMT}{{2003ApJ...589..444G}}

\newcommand{\Ye}{Y_e}

\begin{document}
\title{HARM3D+NUC: A new method for simulating the post-merger phase of binary neutron star mergers with GRMHD, tabulated EOS and neutrino leakage}

\author{Ariadna~Murguia-Berthier}
\affiliation{Department of Astronomy and Astrophysics, University of California, Santa Cruz, CA 95064, USA}
\affiliation{DARK, Niels Bohr Institute, University of Copenhagen, Blegdamsvej 17, 2100 Copenhagen, Denmark}

\author{Scott C. Noble}
\affiliation{Gravitational Astrophysics Lab, NASA Goddard Space Flight Center, Greenbelt, MD 20771, USA}
\author{Luke F. Roberts}
\affiliation{NSCL, Michigan State University, East Lansing, MI 48824, USA}

\author{Enrico~Ramirez-Ruiz}
\affiliation{Department of Astronomy and Astrophysics, University of California, Santa Cruz, CA 95064, USA}
\affiliation{DARK, Niels Bohr Institute, University of Copenhagen, Blegdamsvej 17, 2100 Copenhagen, Denmark}

\author{Leonardo R. Werneck}
\affiliation{Center for Gravitational Waves and Cosmology, West Virginia University, Chestnut Ridge Research Building, Morgantown, WV 26505}

\author{Michael Kolacki}
\affiliation{Center for Computational Relativity, Rochester Institute of Technology, Rochester, New York 14623, USA}
\affiliation{School of Physics and Astronomy, Rochester Institute of Technology, Rochester, New York 14623, USA}

\author{Zachariah B. Etienne}
\affiliation{Department of Physics and Astronomy, West Virginia University, Morgantown, WV 26506}
\affiliation{Center for Gravitational Waves and Cosmology, West Virginia University, Chestnut Ridge Research Building, Morgantown, WV 26505}

\author{Mark Avara}
\affiliation{Center for Computational Relativity and Gravitation, Rochester Institute of Technology, 85 Lomb Memorial Drive, Rochester, New York 14623, USA}

\author{Manuela Campanelli}
\affiliation{Center for Computational Relativity and Gravitation, Rochester Institute of Technology, 85 Lomb Memorial Drive, Rochester, New York 14623, USA}
\affiliation{School of Mathematical Sciences, Rochester Institute of Technology, Rochester, New York 14623, USA}
\affiliation{School of Physics and Astronomy, Rochester Institute of Technology, Rochester, New York 14623, USA}

\author{Riccardo Ciolfi}
\affiliation{INAF, Osservatorio Astronomico di Padova, Vicolo dell'Osservatorio 5, I-35122 Padova, Italy}
\affiliation{INFN, Sezione di Padova, Via Francesco Marzolo 8, I-35131 Padova, Italy}
 
\author{Federico Cipolletta}
\affiliation{Leonardo Corporate LABS - via Raffaele Pieragostini 80, 16149 Genova GE - Italy}

\author{Brendan Drachler}
\affiliation{Center for Computational Relativity and Gravitation, Rochester Institute of Technology, Rochester, New York 14623, USA}
\affiliation{School of Physics and Astronomy, Rochester Institute of Technology, Rochester, New York 14623, USA}

\author{Lorenzo Ennoggi}
\affiliation{Center for Computational Relativity and Gravitation, Rochester Institute of Technology, 85 Lomb Memorial Drive, Rochester, New York 14623, USA}

\author{Joshua Faber}
\affiliation{Center for Computational Relativity and Gravitation, Rochester Institute of Technology, 85 Lomb Memorial Drive, Rochester, New York 14623, USA}
\affiliation{School of Mathematical Sciences, Rochester Institute of Technology, Rochester, New York 14623, USA}
\affiliation{School of Physics and Astronomy, Rochester Institute of Technology, Rochester, New York 14623, USA}

\author{Grace Fiacco}
\affiliation{Center for Computational Relativity and Gravitation, Rochester Institute of Technology, Rochester, New York 14623, USA}
\affiliation{School of Physics and Astronomy, Rochester Institute of Technology, Rochester, New York 14623, USA}

\author{Bruno Giacomazzo}
\affiliation{Universit\'{a} degli Studi di Milano - Bicocca, Dipartimento di Fisica G. Occhialini, Piazza della Scienza 3, I-20126 Milano, Italy}
\affiliation{INFN, Sezione di Milano-Bicocca, Piazza della Scienza 3, I-20126 Milano, Italy}
\affiliation{INAF, Osservatorio Astronomico di Brera, via E. Bianchi 46, I-23807 Merate (LC), Italy}

\author{Tanmayee Gupte}
\affiliation{Center for Computational Relativity and Gravitation, Rochester Institute of Technology, Rochester, New York 14623, USA}
\affiliation{School of Physics and Astronomy, Rochester Institute of Technology, Rochester, New York 14623, USA}

\author{Trung Ha}
\affiliation{Center for Computational Relativity and Gravitation, Rochester Institute of Technology, Rochester, New York 14623, USA}

\author{Bernard J. Kelly}
\affiliation{Department of Physics, University of Maryland Baltimore County, 1000 Hilltop Circle Baltimore, MD 21250, USA}
\affiliation{Gravitational Astrophysics Lab, NASA Goddard Space Flight Center, Greenbelt, MD 20771, USA}
\affiliation{Center for Research and Exploration in Space Science and Technology, NASA Goddard Space Flight Center, Greenbelt, MD 20771, USA}

\author{Julian H. Krolik}
\affiliation{Physics and Astronomy Department, Johns Hopkins University, Baltimore, MD 21218, USA}

\author{Federico G. Lopez Armengol}
\affiliation{Center for Computational Relativity and Gravitation, Rochester Institute of Technology, Rochester, New York 14623, USA}

\author{Ben Margalit}
\affiliation{Astronomy Department and Theoretical Astrophysics Center, University of California, Berkeley, Berkeley, CA 94720, USA}

\author{Tim Moon}
\affiliation{Center for Computational Relativity and Gravitation, Rochester Institute of Technology, 85 Lomb Memorial Drive, Rochester, New York 14623, USA}
\affiliation{School of Mathematical Sciences, Rochester Institute of Technology, Rochester, New York 14623, USA}

\author{Richard O'Shaughnessy}
\affiliation{Center for Computational Relativity and Gravitation, Rochester Institute of Technology, 85 Lomb Memorial Drive, Rochester, New York 14623, USA}
\affiliation{School of Mathematical Sciences, Rochester Institute of Technology, Rochester, New York 14623, USA}
\affiliation{School of Physics and Astronomy, Rochester Institute of Technology, Rochester, New York 14623, USA}

\author{Jes\'{u}s M. Rueda-Becerril}
\affiliation{Center for Computational Relativity and Gravitation, Rochester Institute of Technology, 85 Lomb Memorial Drive, Rochester, New York 14623, USA}

\author{Jeremy Schnittman}
\affiliation{Gravitational Astrophysics Lab, NASA Goddard Space Flight Center, Greenbelt, MD 20771, USA}

\author{Yossef Zenati}
\affiliation{Physics and Astronomy Department, Johns Hopkins University, Baltimore, MD 21218, USA}

\author{Yosef Zlochower}
\affiliation{Center for Computational Relativity and Gravitation, Rochester Institute of Technology, 85 Lomb Memorial Drive, Rochester, New York 14623, USA}
\affiliation{School of Mathematical Sciences, Rochester Institute of Technology, Rochester, New York 14623, USA}
\affiliation{School of Physics and Astronomy, Rochester Institute of Technology, Rochester, New York 14623, USA}

\begin{abstract}
The first binary neutron star merger has already been detected in gravitational waves. The signal was accompanied by an electromagnetic counterpart including a kilonova component powered by the decay of radioactive nuclei, as well as a short $\gamma$-ray burst. In order to understand the radioactively-powered signal, it is necessary to simulate the outflows and their nucleosynthesis from the post-merger disk. Simulating the disk and predicting the composition of the outflows  requires general relativistic magnetohydrodynamical (GRMHD) simulations that include a realistic, finite-temperature equation of state (EOS) and self-consistently calculating the impact of neutrinos. In this work, we detail the implementation of a finite-temperature EOS and the treatment of neutrinos in the GRMHD code HARM3D+NUC, based on \harm.  We include formal tests of both the finite-temperature EOS and the neutrino leakage scheme.  We further test the code by showing that, given conditions similar to those of published remnant disks following neutron star mergers, it reproduces both recombination of free nucleons to a neutron-rich composition and excitation of a thermal wind.
\end{abstract}

\keywords{accretion disks- general relativistic magnetohydrodynamical simulations, neutrino leakage}

\section{Introduction}
On August 17, 2017, the LIGO/VIRGO collaboration detected the first gravitational wave signal arising from the merger of two neutron stars \citep{2017PhRvL.119p1101A}. This signal was accompanied by  a counterpart observed all over the electromagnetic  spectrum \citep{2017ApJ...848L..12A,2017ApJ...848L..34M,Coulter17,2017Sci...358.1574S}. This event, named GW170817, gave credence to the idea that at least a subset of neutron star mergers give rise to short $\gamma$-ray bursts \citep[sGRBs;][]{Eichler1989,1992ApJ...395L..83N,2007NJPh....9...17L,2007PhR...442..166N}. 

In order to understand the electromagnetic emission, we need to study the properties of merger. After the two neutron stars merge, the fate of the remnant depends on the final mass of the resulting object. If the final mass is less than the mass allowed for an object with rigid rotation, then the remnant will be a stable neutron star. On the other hand, if the final mass is larger, then it can result in a hot hyper-massive neutron star (HMNS), supported by differential rotation, or it can promptly collapse to a black hole \citep{2006PhRvD..73f4027S,2008PhRvD..78h4033B,2014MNRAS.441.2433R}. In both cases, the compact object will be surrounded by an accretion disk \citep{Eichler1989,2008PhRvD..78h4033B}.  If the result is an HMNS, there will be transport of mass and angular momentum from the inner edge to the outer edge that will drive the HMNS to rigid rotation, where it can either remain stable, or undergo a delayed collapse to a black hole  \citep[BH; see][for a recent review]{2019arXiv191205659N}. It is widely believed that GW170817 resulted in a delayed collapse to a black hole  \citep{2017ApJ...850L..19M}. In any case, the compact object is left surrounded by an accretion disk containing highly neutron-rich material \citep{2007NJPh....9...17L}.

The post-merger accretion disk will be entirely opaque to photons \citep{1999ApJ...518..356P,2001ApJ...557..949N,2004ApJ...608L...5L,2005ApJ...632..421L,2009ApJ...699L..93L}. As we go deeper in the disk, due to the high density and temperature, neutrinos (and anti-neutrinos) will be created via the charged $\beta$-process, electron-positron annihilation, and plasmon decay \citep{2001ApJ...557..949N,2002ApJ...579..706D,2007ApJ...657..383C}. In the region where the neutrinos are created, matter will be optically thin to neutrinos. In even deeper regions, matter will be optically thick to neutrinos. In the optically thin region,  free neutrinos will carry energy away, and cool the disk, making it {geometrically} thinner \citep{1989ApJ...346..847C,1991ApJ...376..234H}.

Further out in the the disk, where neutrinos are no longer created in substantial numbers, free nucleons will recombine into $\alpha$-particles. The photons will still be trapped in the disk, therefore the disk will be thicker and radiatively inefficient \citep{1999ApJ...518..356P,2001ApJ...557..949N,2004ApJ...608L...5L,2005ApJ...632..421L,2009ApJ...699L..93L}. An outflow arises due to instabilites in the accretion disk from its magnetic field  \citep{1998RvMP...70....1B}. The instabilities will transport angular momentum at significant rates, dissipating energy and driving a high velocity outflow.  In addition, the recombination of free nucleons into $\alpha-$particles is capable of unbinding part of the material from the disk \citep{2009ApJ...699L..93L,2013MNRAS.435..502F}.

Aside from material ejected from the disk, there are other outflows from the binary merger that will significantly contribute to the electromagnetic emission, including a dynamical ejecta \citep[see, for example ][]{1999A&A...341..499R,2015MNRAS.449..390F,2016MNRAS.460.3255R}, and a neutrino-driven wind \citep{dessart09,2013MNRAS.435..502F,Perego14,2017Natur.551...80K,2017CQGra..34o4001F}.  
As the different outflows expand and cool down, heavy elements are synthesised via the rapid neutron capture process ($r$-process) \citep{1999ApJ...525L.121F,2005astro.ph.10256K,2013ApJ...763..108F,2015ApJ...815...82L,2015PhRvD..92d4045P,2016MNRAS.460.3255R,2017MNRAS.464.3907R,2017CQGra..34o4001F,2017MNRAS.472..904L,2018ApJ...869..130R,2019MNRAS.486.1805Z,2020ARNPS..70...95R}. After neutrons are exhausted, elements will radioactively decay and heat the surrounded material, which will thermally emit in the optical/IR bands \citep{1998ApJ...507L..59L,2010MNRAS.402.2771M,2011ApJ...736L..21R, 2013ApJ...774...25K,2013ApJ...775...18B,Tanaka+2013,Grossman+2014,2015MNRAS.450.1777K,2016ApJ...829..110B,2017CQGra..34j4001R,2019ApJ...876..128K,2019EPJA...55..203S}; in particular, see \citet{2019LRR....23....1M} and references within. This emission, called a \emph{kilonova}, was detected for GW170817 \citep{2017Sci...358.1570D,2017Sci...358.1583K,2017ApJ...848L..16S,2017ApJ...848L..27T,2017Natur.551...75S,2017ApJ...848L..18N,2017ApJ...848L..17C,2017ApJ...851L..21V,2017Natur.551...80K,2017Natur.551...67P,2019MNRAS.tmpL..14K}. It is predicted that if the composition of the ejecta includes lanthanides, the emission tends to be more red and peak at later times, whereas if there are no third peak elements, the emission tends to be bluer and peaks earlier \citep{2013ApJ...775...18B,Tanaka+2013}.  Understanding the nucleosynthesis, and the amount of mass ejected is therefore important when deciding the best strategy to observe and perform surveys for kilonovae. This paper will focus on the disk ejecta.

The key parameter in determining the rate of nucleosynthesis, and in particular whether third peak $r$-process elements (including the lanthanides) are created in the disk ejecta, is the electron fraction of the ejected material \citep{2013ApJ...774...25K,2015ApJ...815...82L,2017MNRAS.464.3907R,2017MNRAS.472..904L,2017Natur.551...80K,2021arXiv210208387J}. The problem is that the composition of these ejecta varies between different simulations with results ranging from compositions dominated by iron peak elements to ejecta dominated by lanthanides  \citep[e.g., ][]{2014A&A...568A.105J,2015MNRAS.449..390F,2018PhRvD..98f3007F,2019ApJ...882..163J,2018ApJ...858...52S,2019PhRvD.100b3008M}. One of the significant differences between the simulations is in the neutrino treatment. Neutrinos carry away energy and lepton number, altering the electron fraction and the final ejecta mass and they significantly alter the composition of the ejected material. Thus, simulations need to model the composition and thermodynamic state of the ejecta as realistically as possible to understand and model the kilonova emission.

In order to  model the post-merger disk, we need to self-consistently include multiple relevant physical processes. Due to the compact nature of the BH, we need to consider general relativity (GR). 
Due to the importance of the magnetic stresses we need to include magneto-hydrodynamics (MHD).  
Additionally, to self-consistently include the addition of neutrinos and recombination energy, we need both a realistic equation of state (EOS) and a way in which to consider the impact of neutrinos in the optically thick and thin regions.

There have been many previous efforts to simulate a black hole surrounded by an accretion disk in the context of a binary neutron star. A brief (and certainly incomplete) summary of the numerical efforts is below. 

Numerical simulations initially added neutrino physics by adding pressure terms in the EOS and adding emission and heating/cooling terms from weak reactions in hydrodynamical simulations \citep{1999ApJ...518..356P,2001ApJ...557..949N,2002ApJ...579..706D,2002ApJ...577..311K,2004ApJ...608L...5L,2005ApJ...632..421L,2008MNRAS.390..781M,2011MNRAS.410.2302Z}. 
There have been general relativistic magnetohydrodynamical (GRMHD) simulations in 2d with analytical terms for the neutrino pressure with approximations by \citet{2002ApJ...579..706D} that also include nuclear reactions {using the GRMHD code HARM2D} \citep{2013ApJ...776..105J,2014A&A...568A.105J,2019ApJ...882..163J}.  {There have been} efforts  performing simulations {of binary neutron stars, or a hyper-massive NS, with an accretion disk} in 3d with GRMHD but without neutrinos \citep[for example, ][] {2014ApJ...785L...6S,2014PhRvD..90d1502K,2015PhRvD..92f4034K, 2015PhRvD..92h4064D,2016ApJ...824L...6R,2017PhRvD..95f3016C,2018PhRvD..97l4039K, 2018PhRvD..97b1501R}. Also, groups simulated  disks {after the merger of binary NS} including GR with some kind of neutrino transport  but including no magnetic fields \citep{2016PhRvD..94l3016F,2017ApJ...846..114F,2021ApJ...906...98N}. Other  groups  performed hydrodynamical calculations with neutrino physics, including neutrino leakage schemes and a transport  scheme but no magnetic fields \citep{1996A&A...311..532R,2003MNRAS.342..673R,2014MNRAS.441.3444M,Perego14,2015ApJ...813....2M,2015MNRAS.449..390F,2015MNRAS.453.3386J}.  

\citet{2015PhRvD..91l4021F,2018PhRvD..98f3007F} performed general relativistic hydrodynamical (GRHD) simulations and compared different neutrino treatments, including neutrino transport and leakage schemes. Additionally, \citet{2018ApJ...858...52S} and \citet{2020arXiv201107176D} performed GRMHD simulations of a magnetized torus with a neutrino leakage scheme and the Helmholtz equation of state. { \citet{PhysRevD.97.083014} compared 3d simulations of magnetized and unmagnetized accretion disks with GRMHD  including a neutrino leakage scheme. \citet{2021arXiv210302616L} performed an M1 scheme with neutrino conversions. There have also been GRMHD simulations that included a tabulated EOS with neutrino transport using Monte-Carlo methods  \citep{2019ApJS..241...30M,2019PhRvD.100b3008M}.  }

In this paper, we present simulations using HARM3D+NUC, based on \harm, considering the impact of neutrinos through a leakage scheme and a multi-component, finite-temperature EOS. 
\harm is a versatile GRMHD code that has been well tested and used in many astrophysical scenarios. It uses arbitrary coordinates, allowing for a more accurate conservation of angular momentum. Additionally, it has copious analysis tools developed over the years. The addition of a neutrino leakage scheme and tabulated EOS into HARM3D+NUC is a stepping stone that allows for further advances.
The paper is structured as follows: in Section~\ref{methods} we discuss how we implemented the realistic EOS and the leakage scheme. In Section~\ref{section:validate_eos} we describe the tests we performed to validate the implementation of the tabulated EOS including a torus in hydrostatic equilibrium. In Section~\ref{validate_leakage} we describe the tests we performed to validate the leakage scheme, and in Section~\ref{mhd_torus} we use both the tabulated EOS and leakage scheme to better simulate a torus with a magnetic field.

\section{Methods}
\label{methods}
In order to accurately simulate accretion disks, we need the ability to solve the general relativistic magnetohydrodynamics (GRMHD) equations with a realistic equation of state (EOS) and a way to account for the effect neutrinos and anti-neutrinos have on the material's energy and electron fraction. In this section, we explain how we added a tabulated EOS and neutrino leakage scheme to \harm, in a new code called HARM3D+NUC. 

\subsection{HARM3D+NUC}
\harm \citep{2003ApJ...589..444G,2006ApJ...641..626N, 2009ApJ...692..411N}  solves the GRMHD equations in conservative form.
\harm is a well tested code that can handle arbitrary coordinate systems, which allows for less numerical diffusion and better conservation of angular momentum when using coordinate systems that more closely conform to local symmetries of the problem \citep{2014CQGra..31f5013Z}. Below we set $G=c=1$.
The GRMHD equations of motion include the baryon conservation  equation, 
\beq{
\nabla_\mu \left( n_b u^\mu \right) = 0 \quad , \label{baryon-number-conservation-eq}
}
the energy-momentum conservation equations (with a heating/cooling source, neglecting momentum transfer)
\beq{
\nabla_\mu {T^{\mu}}_\nu  = \mathcal{Q}  u_\nu \quad , \label{energy-conservation-eq}
}
and Maxwell's equations
\beq{
\nabla_\nu \dF^{\mu \nu}  = 0 \quad ,  \label{maxwell-eq-1}
}
\beq{
\nabla_\nu F^{\mu \nu}  = J^\mu \quad ,   \label{maxwell-eq-2}
}
where 
$u^\mu$ is the 4-velocity of the 
fluid, $\mathcal{Q}$ is the energy change rate per
volume in the comoving fluid frame (due to neutrino heating/cooling), $n_b$ is the number density of baryons, $F^{\mu \nu}$ is the Faraday tensor times $1/\sqrt{4\pi}$,
$\dF^{\mu \nu}$ is the dual of this tensor or the Maxwell tensor times 
$1/\sqrt{4\pi}$, and $J^\mu$ is the 4-current\footnote{We follow \cite{\GMT}
in our definition of the electromagnetic field tensor and magnetic field variables.}.
In practice, we don't use Eq.~(\ref{maxwell-eq-2}),  since we work in the limit of ideal MHD.
The change in the conservation of lepton number is
\beq{
\nabla_\mu \left( n_e u^\mu \right) = \mathcal{R}/m_b \quad , \label{lepton-number-conservation-eq}
}
where  $n_e$ is the number
density of electrons, $\mathcal{R}=-\mathcal{R}_{\nu_e}+\mathcal{R}_{{\bar{\nu}_e}}$ is the difference in the net rate of neutrino and anti-neutrino number per
volume in the comoving fluid frame.

Note that the rest-mass density of the gas (mass per unit volume) is dominated by the baryon mass,
$\rho\approx m_b n_b$, where $m_b$ is the baryon mass. 
The baryon number conservation equation can then be replaced by the regular continuity equation:
\beq{
  0 = m_b \nabla_\mu \left( n_b u^\mu \right) = \nabla_\mu \left( m_b n_b u^\mu \right)  = \nabla_\mu \left( \rho u^\mu \right)  \quad .
  \label{continuity-eq}
}
Instead of using $n_e$ and $n_b$, we may use the fluid density $\rho$ and the electron fraction $\Ye$:
\beq{
  \Ye \equiv \frac{n_e}{n_b}  = \frac{n_e}{\rho / m_b} = \frac{m_b n_e}{\rho} 
}
or $\Ye \rho = m_b n_e$ and we can therefore  multiply Eq.~(\ref{lepton-number-conservation-eq}) by $m_b$ to yield the electron
fraction equation:
\beq{
\nabla_\mu \left( \rho \Ye u^\mu \right) = \mathcal{R} \quad . \label{electron-fraction-eq}
}

The total stress-energy tensor is the sum of the fluid part, 
\beq{
T^{\mu\nu}_\mathrm{fluid} = \rho h u^\mu u^\nu 
+ P g^{\mu\nu}  , \label{fluid-stress}
}
and the electromagnetic part 
\beqa{
T^{\mu\nu}_\mathrm{EM} & = & F^{\mu \lambda} {F^\nu}_\lambda 
- \frac{1}{4} g^{\mu \nu} F^{\lambda \kappa}  F_{\lambda \kappa} \\
& = & \bsq u^\mu u^\nu + \frac{1}{2} \bsq g^{\mu\nu} - b^\mu b^\nu \quad 
 , \label{em-stress}
}
where we adopt the ideal MHD condition
\beq{
u_\lambda F^{\lambda \kappa}=0 \quad , 
}
and where
$g_{\mu \nu}$  is the metric, $h =\left(1 + \epsilon + P/\rho \right)$ is
the specific enthalpy, $P$ is the pressure, $\epsilon$ is the 
specific internal energy density, $b^\mu = \dF^{\nu \mu} u_\nu$ 
is the magnetic field 4-vector, and $\bsq \equiv b^\mu b_\mu$ is 
twice the magnetic pressure $P_m$.

Equations~(\ref{energy-conservation-eq}-\ref{continuity-eq}) can be expressed 
in flux conservative form 
\beq{
\del_t \bU\left(\prim\right) = 
-\del_i \bF^i\left(\prim\right) + \mathbf{S}\left(\prim\right) \, 
\label{conservative-eq}
}
where $\bU$ is a vector of ``conserved'' variables, $\bF^i$ are the fluxes, $\mathbf{S}$ is a vector of source terms, and $\prim$ is the vector of primitive variables.  Explicitly, these 
are 
\beq{
\prim=\left[ \rho,  \CTB^k, \tilde{u}^i, \Ye, T\right]^T
}
\beq{
\bU\left(\prim\right) = \sqrt{-g} \left[ \rho u^t , {T^t}_t 
+ \rho u^t , {T^t}_j , B^k , \rho \Ye u^t  \right]^T
\label{cons-U}
}
\beq{
\bF^i\left(\prim\right) = \sqrt{-g} \left[ \rho u^i , {T^i}_t + \rho u^i , {T^i}_j , 
\left(b^i u^k - b^k u^i\right) ,  \rho \Ye u^i  \right]^T
\label{cons-flux}
}
\beq{
\mathbf{S}\left(\prim\right) = \sqrt{-g} 
\left[ 0 , {T^\kappa}_\lambda {\Gamma^\lambda}_{t \kappa} + \mathcal{Q} u_t
, {T^\kappa}_\lambda {\Gamma^\lambda}_{j \kappa} + \mathcal{Q} u_i , 0 , \mathcal{R} \right]^T \, ,
\label{cons-source}
}
where
$g$ is the determinant of the metric, ${\Gamma^\lambda}_{\mu \kappa}$, is
the metric's affine connection, $T$ is the temperature, and $B^i = \CTB^i/\alpha = \dF^{it}$ is the magnetic field.

The primitive velocity is the flow's 4-velocity projected into 
a frame moving orthogonal to the space-like hypersurface:
\beq{
\tilde{u}^\mu = \left({\delta^\mu}_\nu + n^\mu n_\nu \right) u^\nu  }
which  
only has spatial coefficients
\beq{
\tilde{u}^i = u^i + \alpha \gamma g^{ti} \quad ,
 \label{primitive-velocity}
 }
where $\alpha = 1/\sqrt{-g^{tt}}$ is the lapse function, $\beta^i = -g^{ti}/g^{tt}$ is the shift function, 
$\gamma=\alpha u^t$ is the Lorentz factor, and 
$n^\mu$ is the 4-velocity of the orthogonal frame: $n_\mu = [-\alpha,0,0,0]$ and $n^\mu = [1/\alpha, -\beta^i/\alpha]^T$.
Defining a fluid three-velocity $v^i=\tilde{u}^i/\gamma$, it can be shown that $\gamma=1/\sqrt{1-v^2}$, where $v^2=v_iv^i$.

\subsection{Implementation of a tabulated EOS in HARM3D+NUC}
In the following section, we describe the implementation of a tabulated EOS in HARM3D+NUC. 

The tables and routines for interpolating tabulated quantities  are provided by\footnote{The link to the tabulated EOS is the following: {\it https://stellarcollapse.org/SROEOS}, {and the link to the interpolation routines is: {\it https://bitbucket.org/zelmani/eosdrivercxx/src}}} \citet{2010CQGra..27k4103O} and \citet{2017PhRvC..96f5802S}.  The finite-temperature tables give thermodynamic variables, including, for example, the sound speed, and the chemical potentials of the nucleons, electrons/positrons and neutrinos/anti-neutrinos, as a function of the temperature ($T$), the electron fraction ($\Ye$), and the rest-mass density ($\rho$). {The linear interpolation routines are provided by \citet{2010CQGra..27k4103O} and \citet{2017PhRvC..96f5802S}. The interpolation is done in $\log T$, $\log \rho$, $Y_e$ space for $\log \epsilon$, $\log P$, and the rest of the thermodynamical variables.}

The tables consider an interpolation between a single nucleus approximation (SNA) in the high density regime and nuclear statistical equilibrium (NSE) of several nucleides in the low density regime. The SNA is composed of free nucleons, electrons, positrons, $\alpha-$particles, and photons. In the high density regime, nuclei are included using the liquid drop model. The regimes are smoothly interpolated. Using the tables, we have the advantage that the nuclear binding energy release due to recombination energy from the $\alpha$-particles is included.  

There are three main calls to the EOS in HARM3D+NUC:
\begin{itemize}
    \item We call the EOS when setting the characteristic velocity 
    in order to solve the Riemann problem \citep{2003ApJ...589..444G}. The wave velocities depend on the relativistic sound speed \citep{2003ApJ...589..444G}, which can be interpolated directly from the tables. 
    \item We replaced the primitive variable $u=\rho\epsilon$ with the temperature {as a reconstructed variable}, which makes the interpolation of the pressure faster {as all independent variables are known and can be used to perform the interpolation immediately.}. This means that we call the EOS to obtain the primitive energy density $u$ after we update $\rho$, $T$, and $\Ye$ from the conservation equations. 
    \item We call the EOS repeatedly when converting from conserved variables to primitive variables.
\end{itemize}

Our implementation of a tabulated EOS into the conserved to primitive variables routine in HARM3D+NUC follows \citet{2018ApJ...859...71S}.  

\subsubsection{Primary recovery: 3d routine}
The primary recovery routine follows a 3-parameter root-finding method similar to ones implemented in \citet{2008A&A...492..937C,2018ApJ...859...71S}. We call this routine the `3d' routine. For this routine, we reduce the GRMHD equations into three equations that have three unknowns, allowing us to solve the following system:
\beq{
\tilde{Q}^2=\bigg(1-\frac{1}{\gamma^2}\bigg)(\mathcal{B}^2+W)^2-\frac{(Q_\mu\mathcal{B}^\mu)^2(\mathcal{B}^2+2W)}{W^2}
\label{utoprim_solve1}
}
\beq{
Q_\mu n^\mu=-\frac{\CTB^2}{2}\bigg(2-\frac{1}{\gamma^2}\bigg)+\frac{(Q_\mu\CTB^\mu)^2}{2W^2}-W+P(\rho,\Ye,T)
\label{utoprim_solve2}
}
\beq{
\epsilon=\epsilon(\rho, \Ye, T) \quad . 
\label{utoprim_solve3}
}
Using these equations, we perform  Newton-Raphson iterations until we obtain sufficiently accurate values for  the independent variables $\gamma$, {$T$} and $W$.
Here $Q_\mu=-n_\nu T^\nu_\mu=\alpha T^t_\mu$, $W$ is related to the specific enthalpy through $W=h\rho\gamma^2$, 

$\tilde{Q}^\mu={j^\mu}_\nu Q^\nu$, $j_{\mu \nu}=g_{\mu \nu}+n_\mu n_\nu$, and $P$ is the pressure interpolated from tables.

\subsubsection{Backup recovery 1: 2d routine}
We also implemented backup routines that recover the conserved variables. One of them follows an optimized version of the "2d" method of \citet{2006ApJ...641..626N}. We call this routine the `2d' routine. In this routine, the independent variables are $W$ and $v^2$, found using equations~(\ref{utoprim_solve1}-\ref{utoprim_solve2}).  The previous time step's set of primitive variables are used as initial guesses to the Newton-Raphson procedure. As was done in \citet{2018ApJ...859...71S}, we obtain the pressure and the temperature for each $W$ and $v^2$. This is done by first constructing the specific enthalpy: $h=W / (\gamma^2\rho)$, which can also be constructed with quantities from the EOS tables: $h(\rho, T, \Ye)$. Then, with the density, the electron fraction and the specific enthalpy, we perform a Newton-Raphson method to obtain the temperature from the tables, solving the equation: $h=h(\rho, T, \Ye)$. {Note that this inversion is time expensive, which is why this routine is slower than the 3d routine.}

\subsubsection{Backup recovery 2: 2d 'safe-guess' routine}
If there is non-convergence for this backup routine, we include an initial `safe guess' as described in \citet{2008A&A...492..937C}. We call this routine the `2d safe guess' routine. In this scenario, we use the upper limits of the EOS table to obtain the maximum thermodynamical quantities:
\beq{
\rho_{\rm max}=D,
}
\beq{
T_{\rm max}=T_{\rm max,tables} ,
}
\beq{
P_{\rm max}=P(\rho_{\rm max}, \Ye, T_{\rm max}) \quad .
}
Were $D$ is the density measured in the orthogonal frame:
\beq{
D\equiv -\rho n_\mu u^\mu = \gamma \rho \quad . 
}
Then we can estimate the initial `safe guess' for the root-finding procedure:
\beq{
\gamma_{\rm guess}=\gamma_{\rm max}=50,
}
\beq{
W_{\rm guess} =Q_\mu n^\mu+P_{\rm max}-\frac{\CTB^2}{2} \quad .
}
\subsubsection{Backup recovery 3: 2d dog leg routine}
If the 'safe guess' option does not converge, this routine includes a backup root-finding method: a trust-region, dog leg routine that is more robust than a Newton-Raphson \citep{PresTeukVettFlan92,osti_4772677}. We call this routine the `2d dog leg' routine. 
\subsubsection{Backup recovery 4: 'Palenzuela' routine}
If all else fails, we use the routine described in \citet{2015PhRvD..92d4045P}. This routine solves a 1d equation using the Brent method. 
In this routine, called 'Palenzuela', the independent variable is a rescaled variable 
\beq{
x_{\rm pal}\equiv\frac{\rho h\gamma^2}{\rho \gamma} \ .
}
We use the auxiliary rescaled variables:
\beqa{
q_{\rm pal}\equiv\frac{-(Q_\mu n^\mu+D)}{D} & , &\ 
r_{\rm pal}\equiv\frac{\tilde{Q}^2}{D^2} \ ,\ \\
s_{\rm pal}\equiv\frac{\mathcal{B}^2}{D} & , & \ 
t_{\rm pal}\equiv\frac{Q_\mu\CTB^\mu}{D^{3/2}} \ .
}
The independent variable should be bracketed between:
\beq{
1+q_{\rm pal}-s_{\rm pal}>x_{\rm pal}>2+2q_{\rm pal}-s_{\rm pal} \ .
}
The method uses an initial guess for $x_{\rm pal}$ from the previous time step, and gets approximate quantities. Using them, it updates $x_{\rm pal}$ and iterates again until convergence is reached.
The method is the following (where approximate quantities will be denoted by a hat):

We obtain an approximate Lorentz factor $\hat{\gamma}^{-2}$:
\beq{
\hat{\gamma}^{-2}=1-\frac{x_{\rm pal}^2r_{\rm pal}+(2x_{\rm pal}+s_{\rm pal})t_{\rm pal}^2}{x_{\rm pal}^2(x_{\rm pal}+s_{\rm pal})^2} \ .
}
With that, we can estimate:
\beq{
\hat{\rho}=\frac{D}{\hat{\gamma}}
}
and an approximate specific energy:
\beq{
\hat{\epsilon}=\hat{\gamma}-1+\frac{x_{\rm pal}}{\hat{\gamma}}(1-\hat{\gamma}^2)+\hat{\gamma}\bigg{(}q_{\rm pal}-s_{\rm pal}+\frac{t_{\rm pal}^2}{2x_{\rm pal}^2}+\frac{s_{\rm pal}}{2\hat{\gamma}^2}\bigg{)} \ .
}
A call to the EOS will give the pressure $\hat{P}(\hat{\rho}, \hat{\epsilon}, Y_e)$, and with all those approximate quantities, we can solve for $x_{\rm pal}$ using the Brent method by solving:
\beq{
0=f(x_{\rm pal})=x_{\rm pal}-\hat{\gamma}\bigg{(} 1+\hat{\epsilon}+ \frac{\hat{P}}{\hat{\rho}}\bigg{)} \ .
}
We repeat the estimation of all the hat quantities until the solution for $x_{\rm pal}$ converges.

\begin{figure*}[t!]
\includegraphics[scale=0.275]{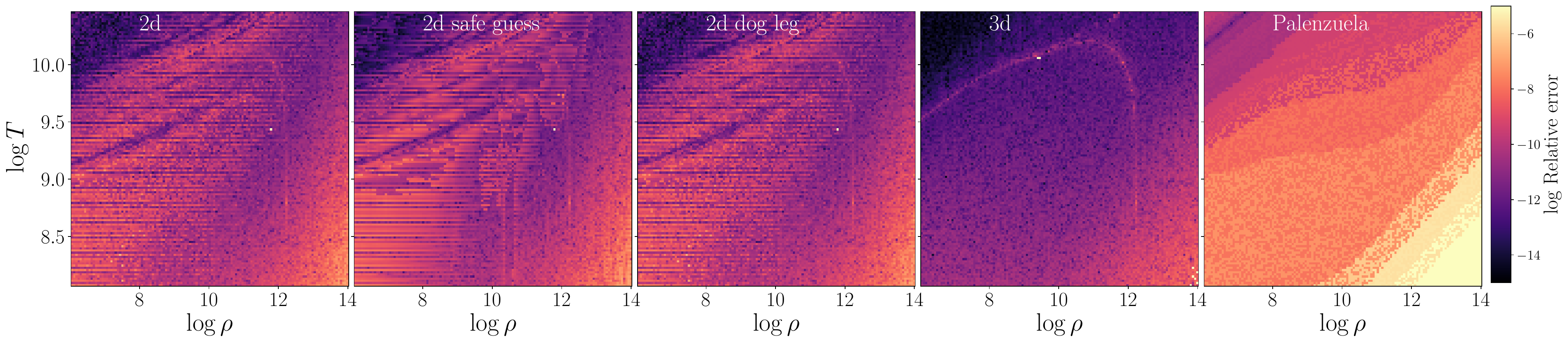}

\caption{Relative error comparing primitive variables created from a grid of density and temperature after we performed the conversion from conserved variables to primitive variables. The primitive variables were created with $\Ye=0.1$, the Lorentz factor $\gamma=2$, $\log{\frac{P_{\rm mag}}{P_{\rm gas}}}=-5$, and a Minkowski metric. We perturbed them by $5\%$ and then recovered them using our conserved to primitive routines. The error is calculated by summing over the relative error of each primitive variable compared to the original. We did this for $2^{14}$ points in the shown range. The 2d routines failed only once,  the 3d routines failed 11 times, and the Palenzuela routine did not fail in this range. Density is in units of ${\rm g/cm^3}$, and temperature is in units of Kelvin. Here we compare different routines, described in the text.}
\label{fig:error_plot}
\end{figure*}

\subsection{Neutrino leakage scheme}
In the following section, we describe how we implemented a leakage scheme that takes into account the  heating/cooling due to neutrinos, as well as how their emission and absorption affect the electron fraction. This leakage scheme is suited to describe the contribution of neutrinos to the composition, and energy.

\subsubsection{Rates}
The scheme calculates the absorption/emission rate as well as the energy loss rates due to neutrinos. We use these rates in the source terms of Eq.~(\ref{energy-conservation-eq}) and Eq.~(\ref{lepton-number-conservation-eq}). The scheme uses energy-averaged quantities.

Like \citet{1996A&A...311..532R,2013PhRvD..88f4009G,2018ApJ...858...52S}, we consider the following neutrino reactions, each with their own absorption/emission rate (which has units of $\rm{cm}^{-3}{\rm s}^{-1}$) and the energy loss rate rate due to neutrinos (with units of $\rm{erg}\, \rm{cm}^{-3}{\rm s}^{-1}$):

\begin{itemize}
\item Charged $\beta$-process with  $\mathcal{R}^\beta_{\nu_i}$ and $\mathcal{Q}^\beta_{\nu_i}$:
\beq{
e^-+p\rightarrow n+\nu_e }
\beq{
e^+ +n\rightarrow p+\bar{\nu}_e
}
\item{Plasmon decay with  $\mathcal{R}^{\gamma}_{\nu_i}$ and $\mathcal{Q}^\gamma_{\nu_i}$:
 \beq{
 \gamma \rightarrow \nu_e+ \bar{\nu}_e}
\beq{
  \gamma \rightarrow \nu_x+ \bar{\nu}_x
 }
 where x is the muon and tauon, and in this case, $\gamma$ corresponds to a photon.}
 
\item Electron-positron pair annihilation with  $\mathcal{R}^{ee}_{\nu_i}$ and $\mathcal{Q}^{ee}_{\nu_i}$
\beq{
e^-+e^+\rightarrow \nu_e+\bar{\nu}_e}
 \beq{
e^-+e^+\rightarrow \nu_x+\bar{\nu}_x \ . } 
\end{itemize}

Using the above reactions, we calculate the total number emission in the optically thin regime from species $i$ as \citep{1996A&A...311..532R}:
\beq{\mathcal{R}_{\nu_i}=\mathcal{R}^\beta_{\nu_i}+\mathcal{R}^{\gamma}_{\nu_i}+\mathcal{R}^{ee}_{\nu_i}
\label{eq:emission_rate}
}
and the total energy loss rate rate in the optically thin regime is:
\beq{\mathcal{Q}_{\nu_i}=\mathcal{Q}^\beta_{\nu_i}+\mathcal{Q}^{\gamma}_{\nu_i}+\mathcal{Q}^{ee}_{\nu_i} \quad , 
\label{eq:heating_rate}
}
where "$i$" denotes the different neutrino/anti-neutrino flavors: electron, or muon and tauon.

The total emission/absorption rates and the energy loss rates are given by an interpolation between the diffusive optically thick regime and the transparent optically thin regime \citep{1996A&A...311..532R}:
\beq{
\mathcal{R}_{\nu_i}^{\rm eff}=\mathcal{R}_{\nu_i}\left(1+\frac{t_{\rm diff}}{t_{\rm emission, \mathcal{R}}}\right)^{-1}
}
\beq{
\mathcal{Q}_{\nu_i}^{\rm eff}=\mathcal{Q}_{\nu_i}\bigg{(}{1+\frac{t_{\rm diff}}{t_{\rm emission, \mathcal{Q}}}}\bigg{)}^{-1}. \ .
}
Here the diffusion timescale is given by:
\beq{t_{\rm diff}= \frac{D_{\rm diff} \tau^2}{c\kappa_{\nu_i}} \ ,
}
where $D_{\rm diff}=6$ \citep{2003MNRAS.342..673R,2010CQGra..27k4103O,2018ApJ...858...52S} and $\tau$ is the optical depth, and $\kappa_{\nu_i}$ the energy averaged opacity (in units of $\mathrm{cm}^{-1}$) of $\nu_i$. The absorption/emission and energy loss timescales are $t_{\rm emission, \mathcal{R}}=\mathcal{R}_{\nu_i}/n_{\nu_i}$, with $n_{\nu_i}$ being the neutrino number density (at chemical equilibrium), and $t_{\rm emission, \mathcal{Q}}=\mathcal{Q}_{\nu_i}/\varepsilon_{\nu_i}$, with $\varepsilon_{\nu_i}$ being the neutrino energy density. In the optically thick regime, the neutrino loss rate is less than the diffusion time, which results in $\mathcal{R}_{\nu_i}^{\rm eff}=n_{\nu_i}/t_{\rm diff}$ and $\mathcal{Q}_{\nu_i}^{\rm eff}=\varepsilon_{\nu_i}/t_{\rm diff}$, whereas in the optically thin regime, we recover the rates from equation~(\ref{eq:emission_rate}) and~(\ref{eq:heating_rate}). The rates for the muon and tauon neutrinos/anti-neutrinos estimated in \citet{1996A&A...311..532R}   take into account all four of those species. We also note that several quantities, including the chemical potentials, are obtained from EOS table interpolation.

\subsubsection{Optical depth}
The transition between the two regimes will be set by the optical depth $\tau_{\nu_i}$, which is also needed to obtain the diffusion timescale. In order to get the optical depth, we consider the following reactions as the source of neutrino opacity:

\beq{
\nu_e+n\rightarrow p+e^-
}
\beq{
\bar{\nu}_e+p\rightarrow n+e^+ \ 
}
\beq{
\nu_i + p\rightarrow\nu_i + p
}
\beq{
\nu_i + n\rightarrow\nu_i + n.
}

The opacities are obtained from \citet{1996A&A...311..532R}.  Electron scattering is neglected.

{The usual global approach to calculate  the optical depth of a point in the flow would be to integrate the opacity over all directions and determine the path of minimal absorption. This approach assumes that the neutrino will follow a straight path. However, we follow \citet{2014PhRvD..89j4029N,2018ApJ...858...52S}, where a local, iterative approach is used instead of a global calculation, and where crooked minimal paths are acceptable. The optical depth is calculated by obtaining the shortest path of the neutrino out of the star using its neighbors. For the first timestep, we begin by initializing the optical depth grid to zero. Next, we perform the first iteration, where we estimate the optical depth at each cell as the minimum of the optical depth of its neighbor ($\tau_{{\nu_i}, \rm{neighbor}}$) plus the optical depth needed for the neutrino to reach that neighbor ($\bar{\kappa}_{\nu_i} {(g_{{k}j}dx^{{k}}  dx^j)}^{1/2}$):}
\beq{
\tau_{\nu_i}=\min\big{(}\tau_{{\nu_i}, \rm{neighbor}}+\bar{\kappa}_{\nu_i} {(g_{{k}j}dx^{{k}}  dx^j)}^{1/2}\big{)}
}
where $\tau_{{\nu_i}, \rm{neighbor}}$ is the optical depth of the neighboring cell, $\bar{\kappa}_{\nu_i}$ is the average opacity between the cell and its neighbor, and  ${(g_{{k}j}dx^{{k}} dx^j)}^{1/2}$ is the  distance to the neighboring cell calculated by taking the average value of $g_{kj}$ between the local and neighboring cells. {We minimize over all neighbors.}

{This essentially traces the path of least resistance of the neutrino to a neighbor. We update the entire grid, and perform the next iteration, where again, we minimize over all the adjacent neighbors. 
The next iteration will show the path to the neighbor two cells away. As we do more iterations, we trace the path of least resistance that the neutrinos will take out of the star. This will lead us to the final optical depth.}
{During the first timestep, we initialize the optical depth by by performing $20N_{\rm max}$ iterations, where $N_{\rm max}$ is the maximum number of cells in each direction, independent of resolution. This is done to trace a path to the edge of the domain initially.}
After the initial calculation, which has a fixed number of iterations, {we continue to do iterations to obtain the final optical depth, however} we impose a convergence criterion in order to minimize the number of iterations. In order to converge, we set conditions on the difference between iteration ${k}-1$ and ${k}$:
\beq{
R_{\rm change,\tau}({k}) \equiv \frac{| \sum \tau_{{k}-1}-\sum \tau_{{k}}|}{\sum \tau_{{k}-1}}<\epsilon_1 
}
or 
\beq{
\frac{| R_{\rm change,\tau}({k}-1)-R_{\rm change,\tau}({k})|}{R_{\rm change,\tau}({k}-1)}<\epsilon_2 
}
where $\sum \tau_k$ is the sum of all the optical depths in the grid at iteration ${k}$, and $\epsilon_1$  and $\epsilon_2$ are parameters that we choose to be $\epsilon_1=10^{-4}$ and $\epsilon_2=10^{-3}$, respectively.
Only a few iterations are needed for convergence after the initial guess.

\section{Validation tests for the tabulated EOS}
\label{section:validate_eos}
In this section, we describe the tests performed to validate the implemented EOS tables.

\subsection{Testing the conserved to primitive variables routine}
In order to validate the routines that transform the conserved variables into primitive variables with tabulated EOS, we created primitive variables out of a grid of density and temperature values within the EOS table. The magnetic field was set randomly to  be either aligned or anti-aligned with the velocity vector. The magnitude of the magnetic field was set to be such that: $b^2/2=\left(\frac{P_{\rm mag}}{P_{\rm gas}}\right) P_{\rm gas}$, where $(P_{\rm mag}/P_{\rm gas})$ is set as a parameter, $P_{\rm mag}$ is the magnetic pressure, and  $P_{\rm gas}$ is the gas pressure.
We then obtained a set of conserved variables based on these primitives. 
The true primitives were then varied  by  randomly  adding or subtracting a $5\%$ perturbation to each primitive. This test is based on \citet{2018ApJ...859...71S}.

We then used these primitives as initial guesses for the various routines that transform the conserved variables to primitive variables and compared the resultant solution to the original.

We show the error we obtained for all primitive variables in Figure~\ref{fig:error_plot}. 
It can be seen that the recovery error is low. Additionally, the figure shows that the 3d method is less robust, but more accurate, which is the reason it is set as the primary routine. The different 2d methods, and the 'Palenenzuela' routine  are more robust, but less accurate (and slower) than the 3d method, so they serve better as backup routines. 

\begin{figure}[t!]
\includegraphics[scale=0.24]{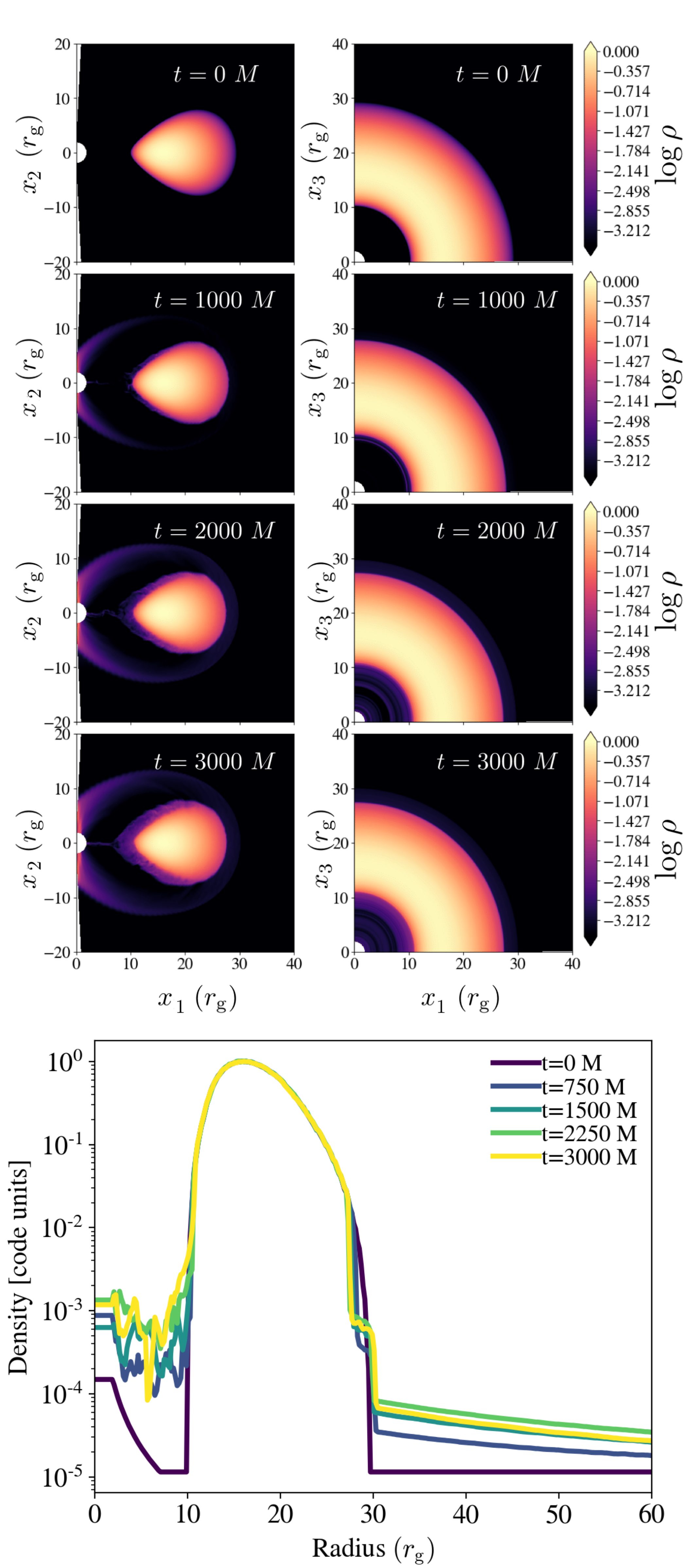}
\caption{\textit{ Top} panel: Evolution of a torus in hydrostatic equilibrium {with a tabulated EOS and no neutrino leakage scheme.}. We show the meridional (Left) and equatorial (Right) cut. The initial conditions are set as described in section~\ref{torus_initial_conditions}. Here $x_1$, $x_2$, $x_3$  correspond to the coordinates $x$, $z$, $y$ respectively. {\textit{Bottom} panel: Density as a function of radius for different times in the equator.}}
\label{fig:hydro_torus}
\end{figure}

\begin{figure}[t!]
\includegraphics[scale=0.57]{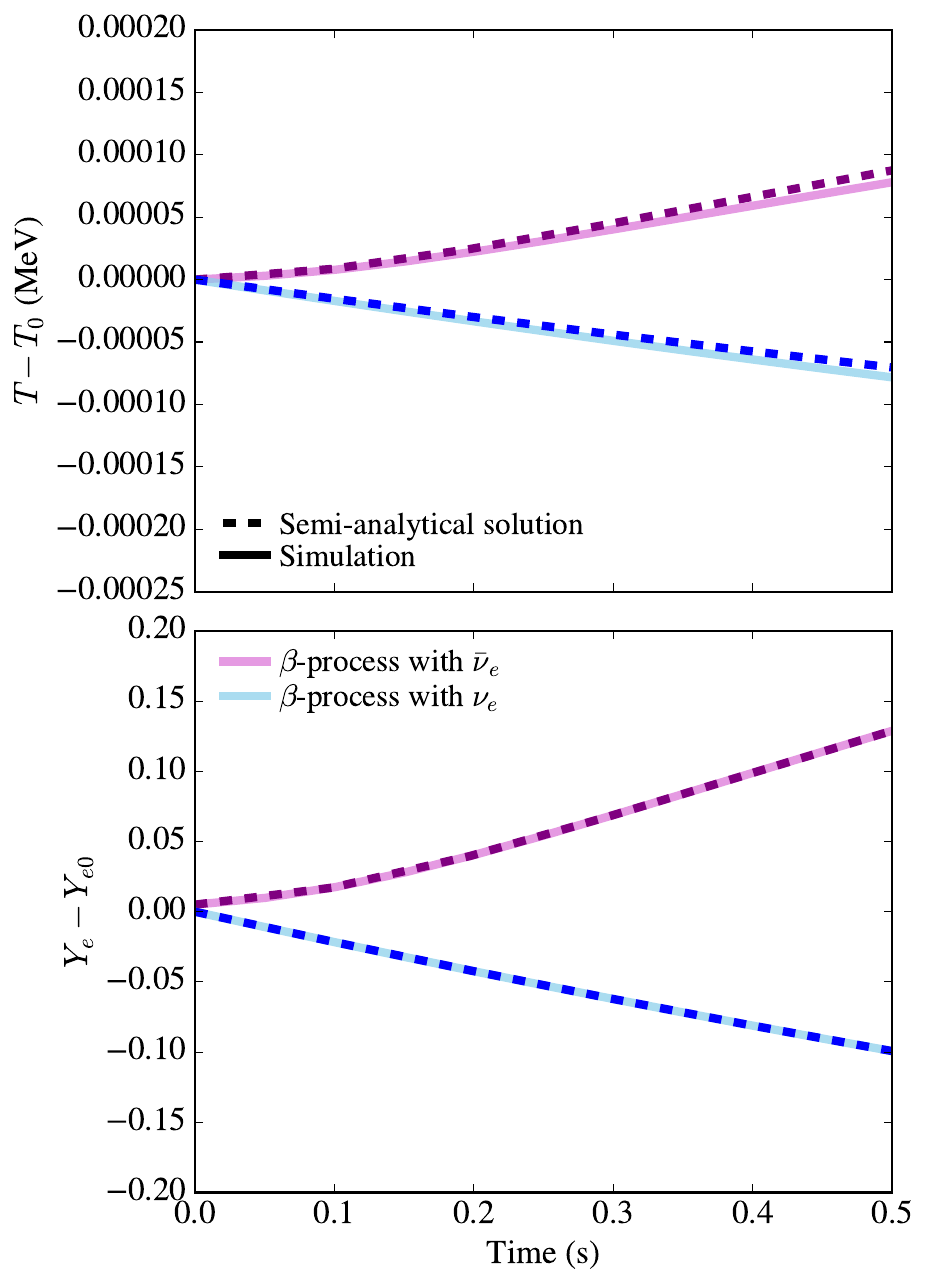}
\caption{Comparison of the semi-analytical solution (dotted line) with the simulation (solid line) for the evolution of the temperature and electron fraction of an isotropic, optically thin gas with constant density. }
\label{fig:beta_proc}
\end{figure}

\begin{figure}[t!]
\includegraphics[scale=0.6]{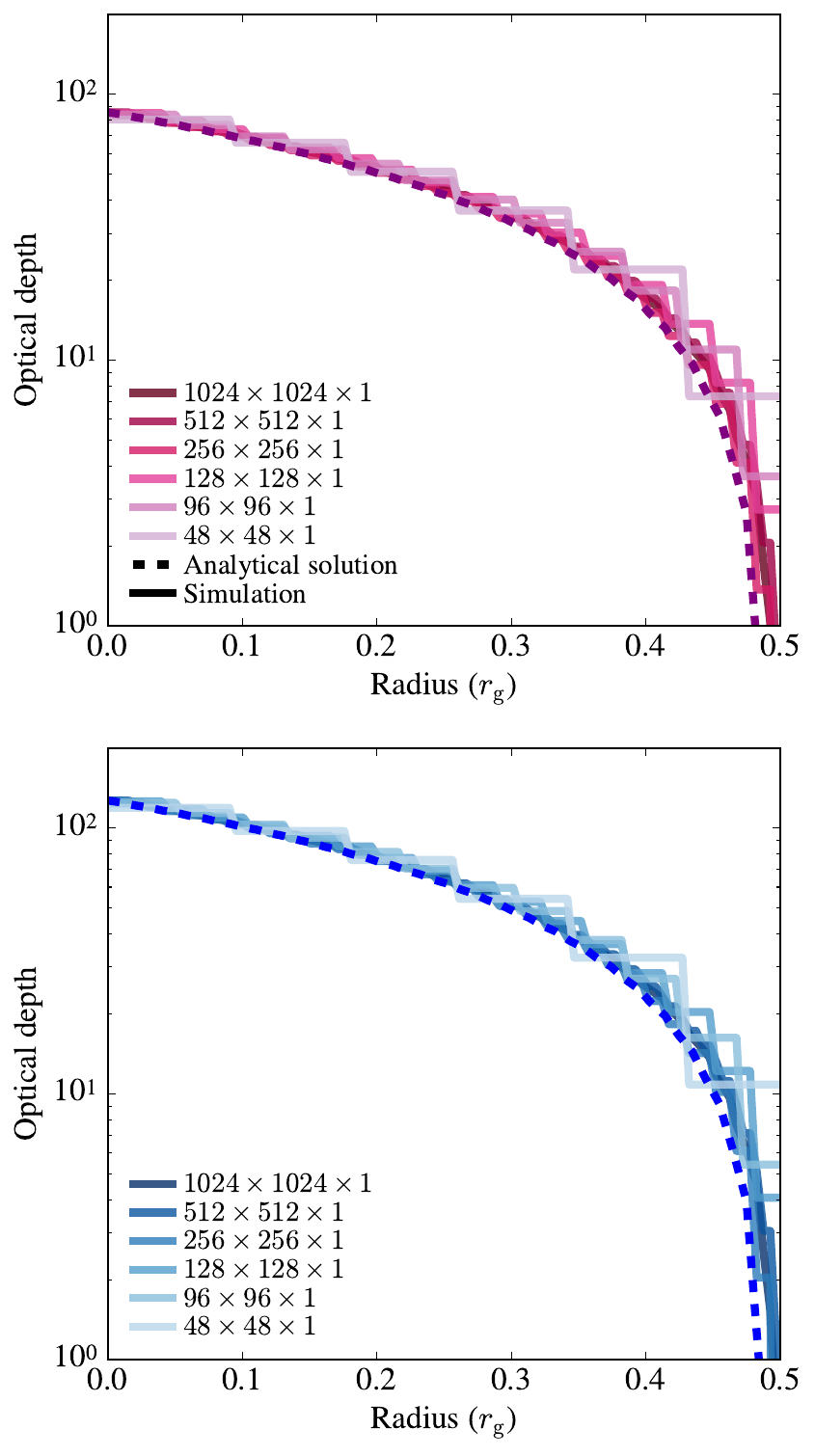}
\caption{Comparison of the analytical solution (dotted line) of the optical depth with simulations (solid line) for different resolutions. The labels indicate the number of cells in each direction.  The {\it Top} panel is the anti-neutrino optical depth, and the {\it Bottom} panel is the neutrino optical depth.  }
\label{fig:optical_depth}
\end{figure}

\begin{figure}[t!]
\includegraphics[scale=0.6]{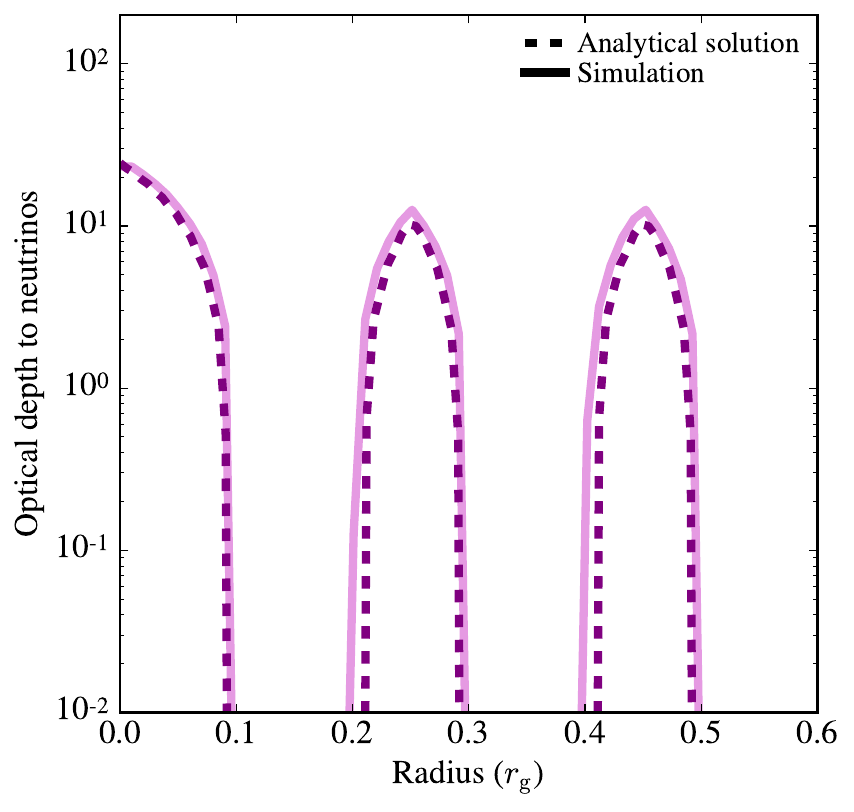}
\caption{Comparison of the analytical solution (dotted line) of the optical depth to neutrinos with simulations (solid line).   }
\label{fig:stripes}
\end{figure}

\subsection{Torus in hydrostatic equilibrium}
\label{torus_initial_conditions}
To test the EOS implementation, we simulated a non-magnetized torus that is in hydrostatic equilibrium with no leakage scheme, following \citet{1976ApJ...207..962F}.

Figure~\ref{fig:hydro_torus} shows the 3d hydrodynamical evolution of a torus constructed to be in hydrostatic equilibrium with a tabulated EOS {without neutrino cooling}. There are perturbations particularly near the BH due to accretion onto the BH, but the density is low in those regions. As can be seen from the figure, the torus remains in hydrostatic equilibrium throughout the simulation.

\subsubsection{Initial conditions inside the torus}
\label{initial_hydro}
The specific enthalpy inside the torus is implemented via Equation~(3.6) of \citet{1976ApJ...207..962F}, but adding  $\ln{h_{\rm min}}$ to the integration constant (see~Section~\ref{atmosphere_torus}).
By construction, the torus is in hydrostatic equilibrium with the ambient atmosphere. We also set the torus to be isentropic, and have uniform electron fraction. Given a specific entropy {$s_{\rm disk}$}, a specific enthalpy {given by \citet{1976ApJ...207..962F}}, and an electron fraction $Y_{e\rm, disk})$, the temperature and density of the disk are found by solving the following equations:
\beq{
s_{\rm disk}=s(\rho, T, \Ye)
}
\beq{
h=h(\rho, T, \Ye)
}
where $s$ is the specific entropy.

\subsubsection{Atmosphere}
\label{atmosphere_torus}
In the classical torus, the boundaries of the torus are defined where $h=1$. In the tabulated EOS, though, negative internal energy densities are allowed since the internal energy per nucleon is measured relative to the free neutron rest mass energy. In this case, the minimum specific enthalpy is not restricted to $1$, but rather it can be $1>h_{\rm min}>0$, where $h_{\rm min}$ is the specific enthalpy from the table given the atmospheric density, and the disk's electron fraction and specific entropy. Thus, we set the torus boundary to be where $h=h_{\rm min}$.  
{For the background atmosphere, we set the minimum  atmospheric density $\rho_{\rm atm}$ as a parameter. Then we find the minimum  specific enthalpy by doing a table inversion and finding $h_{\rm min}=h(\rho_{\rm atm},s_{\rm disk},Y_{e\rm, disk})$. We also find the atmospheric temperature by doing a table inversion $T_{\rm atm}=T(\rho_{\rm atm},s_{\rm disk},Y_{e\rm, disk})$.}

The density in the background is set to:
\beq{
\max\bigg{(}\rho_{\rm atm},\frac{\rho_0}{r^2}\bigg{)}}
Where we set $\rho_0$ as a parameter as well. The background atmosphere temperature is set to:
\beq{
\max\bigg{(}T_{\rm atm}, \frac{T_0}{r}\bigg{)} \ , }  
where $T_0$ is a parameter. 
The power-law dependence is set to provide the background atmosphere with more pressure support so that it does not rapidly accrete onto the BH. This ultimately helps with robustness near the BH, as the low density and low temperature zones with high velocity are where the conserved to primitive routines tend to fail. We note that in the region where there is a power-law dependence, {the specific enthalpy is not a constant}, whereas once the background atmosphere is set to be constant, everything is thermodynamically consistent because it was constructed with the tabulated EOS tables.
We set the electron fraction of the atmosphere to a constant value  found by assuming $\beta$-equilibrium (where the neutrino chemical potential is zero) at $T_{\rm atm}$ and $\rho_{\rm atm}$.

The units are normalized so that the maximum density in the torus is set to $\rho_{\rm max}=1$ {in code units, which in this case corresponds to $\rho_{\rm max}=5.4\times10^8{\rm g/cm^3}$ in cgs units}. In the simulation we performed, the torus has a constant electron fraction of $\Ye=0.1$ and a specific entropy of $10{\rm k_B/baryon}$, where ${\rm K_B}$ is Boltzmann's constant. The background atmosphere is characterized by $\rho_{\rm atm}=6000 {\rm g/cm^3}$, $\rho_{0}=3 \times 10^5 {\rm g/cm^3}$, $T_{0}=0.4$MeV.  We used the SLy4 table with NSE from \citet{2017PhRvC..96f5802S}, and with that table the minimum specific enthalpy for our parameters is set to $h_{\rm min}=0.9974$ (in code units), and $T_{\rm atm}=0.0053$MeV. The electron fraction in the atmosphere, given by $\beta$-equilibrium, is set to $Y_{e,{\rm atm}}=0.45$. The boundary conditions are outflow in the outer radial boundary, reflective in the angular coordinate $\theta$, and periodic in the angular coordinate $\phi$. The metric is Kerr-Schild in spherical coordinates for a non-spinning BH.

\section{Validation tests for the leakage scheme}
\label{validate_leakage}
In this subsection, we describe how we tested the leakage scheme in the optically thin regime and for finite optical depth.

\subsection{Testing the optically thin regime}

Following \citet{2019ApJS..241...30M}, we tested the leakage scheme in an optically thin regime by considering an isotropic gas of constant density and temperature such that the gas is optically thin to neutrinos. We tested both reactions in the charged $\beta-$process separately where we included only either the neutrinos or the anti-neutrinos.

In this case, the GRMHD equations reduce to:

\beq{
\partial_t{{T^t}_t}=\mathcal{Q}. \ ,
}
\beq{
\partial_t{Y_{e}}=\mathcal{R}/\rho \ , 
}
where $\mathcal{R}$ and $\mathcal{Q}$ are the 
emission/absorption and energy loss rates due to neutrinos or anti-neutrinos of the reactions in $\beta-$process separately. The rates need to be calculated semi-analytically, since they depend on interpolated quantities, such as the degeneracy parameters. We can then solve the equations semi-analytically with a set of initial conditions and compare to simulations.  We chose the initial density and temperature such that the medium is optically thin to neutrinos and anti-neutrinos.

For the initial conditions, we used an initial density of $617714 {\rm g/cm^3}$ and temperature of $1$MeV, chosen so that the medium is optically thin to neutrinos and anti-neutrinos. We used $Y_{e,0}=0.5$, $Y_{e,0}=0.005$ for the electron neutrino and anti-neutrino tests respectively. We used $2\times2\times1$ number of cells in each direction using a Cartesian grid with Minkowski metric.

In Figure~\ref{fig:beta_proc} we show the comparison between the semi-analytical solution and the simulation for the $\beta$-process both for neutrinos and anti-neutrinos. We compare the change in the electron fraction due to the absorption/emission rate, and the change in temperature due to the heating/cooling rate. As can be seen from the figure, HARM3D+NUC is able to recreate the semi-analytical solution.

\subsection{Testing the optically thick regime}
\subsubsection{Constant density circular disk}
In order to test the optical depth calculation, we simulated a circular disk with uniform density and  temperature embedded in an optically thin medium of constant density and temperature. The advantage of this scenario is that we can calculate the opacity inside the circle and then calculate the optical depth analytically. This way we can compare to the simulation. 
The simulations were performed in 2d, and the domain is $2r_{\rm g}$, where $r_{\rm g}=G M / c^2$ is the gravitational radius. We used a Minkowski metric with spherical coordinates. There are outflow conditions on the radial boundaries.
The optical depth in the outer radial boundary was set to zero so that the neutrinos and anti-neutrinos could escape the domain. We simulated an optically thick circular disk that has a constant density of $9.8\times 10^{13}{\rm g/cm^3}$, an electron fraction of $0.1$ and a temperature of $8$MeV embedded in an optically thin medium, with a density of $6\times 10^7{\rm g/cm^3}$, an electron fraction of $0.5$ and a temperature of $0.01$MeV.
Figure~\ref{fig:optical_depth} shows the optical depth for both the electron neutrino and anti-neutrino for different resolutions.  As can be seen from the figure, the initial guess for the optical depth is accurate and the convergence to the solution does not change with resolution. At smaller optical depths, the optical depth is slightly overestimated at lower resolutions, but as the optical depth increases, the solution doesn't depend noticeably on resolution.

\subsubsection{Stripes}
We can also test the optical depth algorithm by simulating stripes of high density material with low density material in between. In this scenario, it is expected that a neutrino created in the  region with high optical depth material will travel to the region with low optical depth and stream freely from the surface. For the simulation, we used $4096\times96\times 1$ cells. The simulations were performed in 2d, and the domain is $1r_{\rm g}$ large in radial extent, where $r_{\rm g}=G M / c^2$ is the gravitational radius. We used a Minkowski metric with spherical coordinates and outflow conditions at the radial boundaries.
The optical depth at the outer radial boundary was set to zero so that the neutrinos and anti-neutrinos could escape the domain.  We simulated three stripes of material with high optical depth:  $\rho = 9.8\times 10^{13}{\rm g/cm^3}$, $Y_e = 0.1$, $T = 8\mathrm{MeV}$. In between the stripes, the optically thin gas was initialized to $\rho=6\times 10^7\mathrm{ g/cm^3}$, $Y_e = 0.5$, and $T=0.01\mathrm{MeV}$. The high opacity stripes start at $r=0 r_{\rm g}$, and have a width of $r=0.1 r_{\rm g}$. The next stripes are located in $r=0.2r_{\rm g}$ and $r=0.4 r_{\rm g}$.

We show the results from this setup in Figure~\ref{fig:stripes}, where we compare the results from the simulation with the analytical estimate (length units are in $r_{\rm g}$):
\beq{
\tau_{\rm analytical}=
        \left\{ \begin{array}{lllll}
             \int_{0}^{0.2} \kappa dr  & r \leq 0.2 \\
            \int_{0.2}^{0.25} \kappa dr  & 0.2 \leq r \leq 0.25 \\
            \int_{0.25}^{0.35} \kappa dr  & 0.25 \leq r \leq 0.35 \\
            \int_{0.4}^{0.45} \kappa dr  & 0.4 \leq r \leq 0.45 \\
            \int_{0.45}^{0.55} \kappa dr  & 0.45 \leq r \leq 0.55
        \end{array} \right.
}

\begin{figure}[t!]
\includegraphics[scale=0.23]{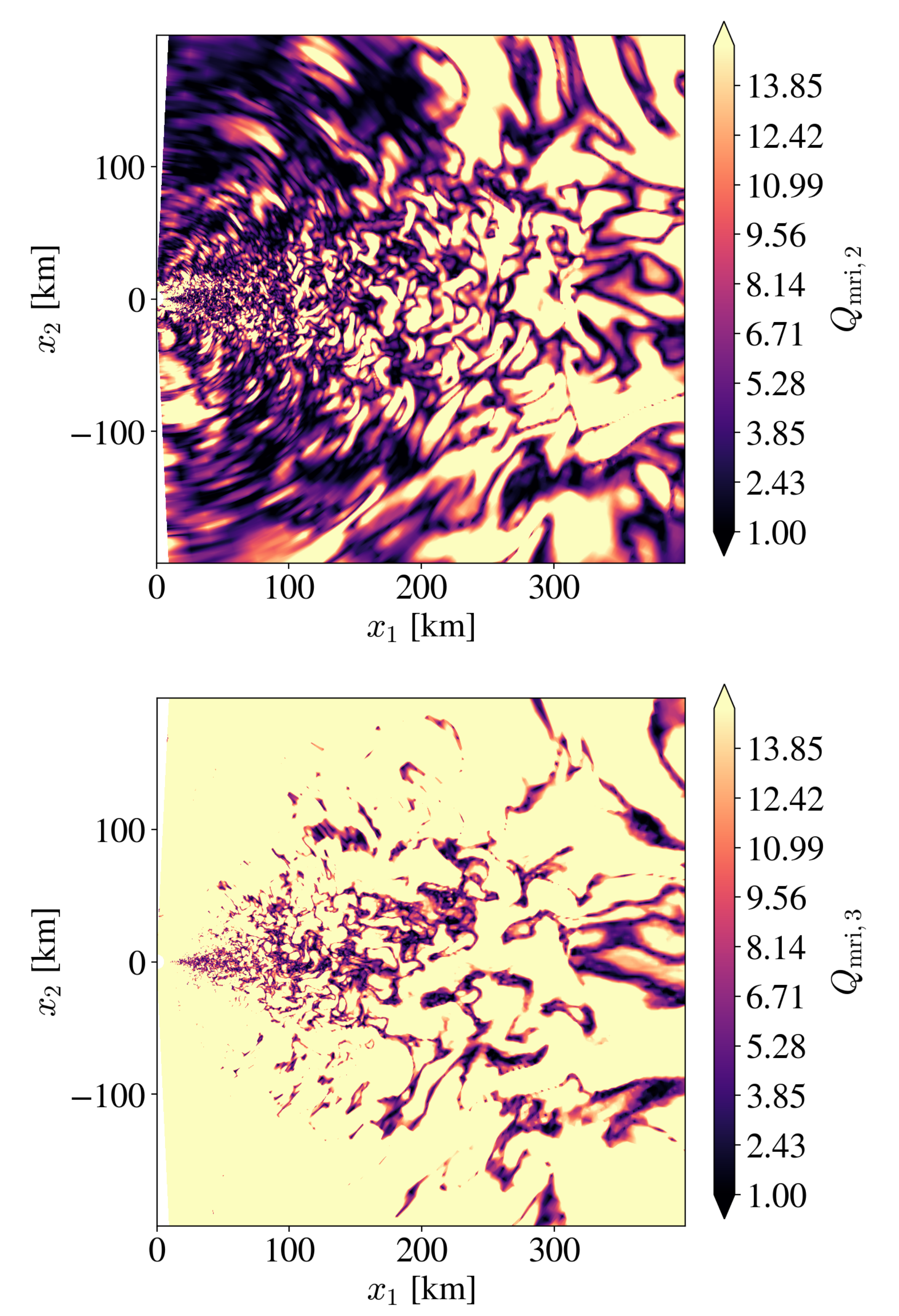}
\caption{Shown is a meridional cut of the MRI quality factors $Q_{ \rm mri, 2}$ and $Q_{\rm mri, 3}$ at {114}ms, where the subscripts for $Q_{\rm mri,2}$, $Q_{\rm mri,3}$ correspond to the coordinates $\theta, \phi$ respectively.}
\label{fig:mhd_q_mri_colormap}
\end{figure}

\begin{figure}[t!]
\includegraphics[scale=0.23]{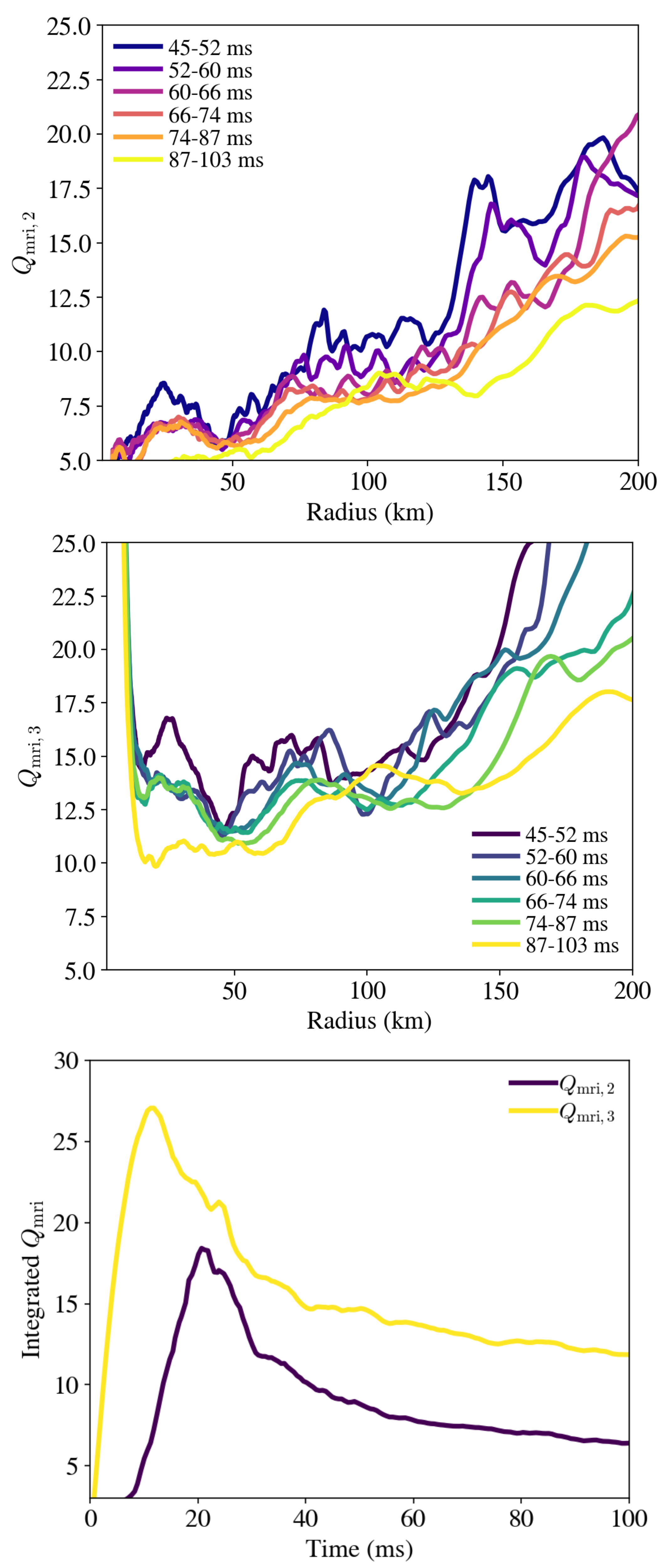}
\caption{ {\textit{Top} and \textit{Middle} panel: Shell-integrated mass-weighted quality factors  as a function of radius, averaged over different epochs of time.
\textit{Bottom} panel: Mass-weighted quality factors integrated over angles and radii that are less than 150km: $\int^{150\rm km}_{0\rm km}\int \int Q_{\rm mri}\rho \sqrt{-g} dr  d\phi d\theta/\int^{150\rm km}_{0\rm km}\int \int  \sqrt{-g}\rho dr d\phi d\theta$. }}
\label{fig:mhd_q_mri}
\end{figure}

\begin{figure}[t!]
\includegraphics[scale=0.23]{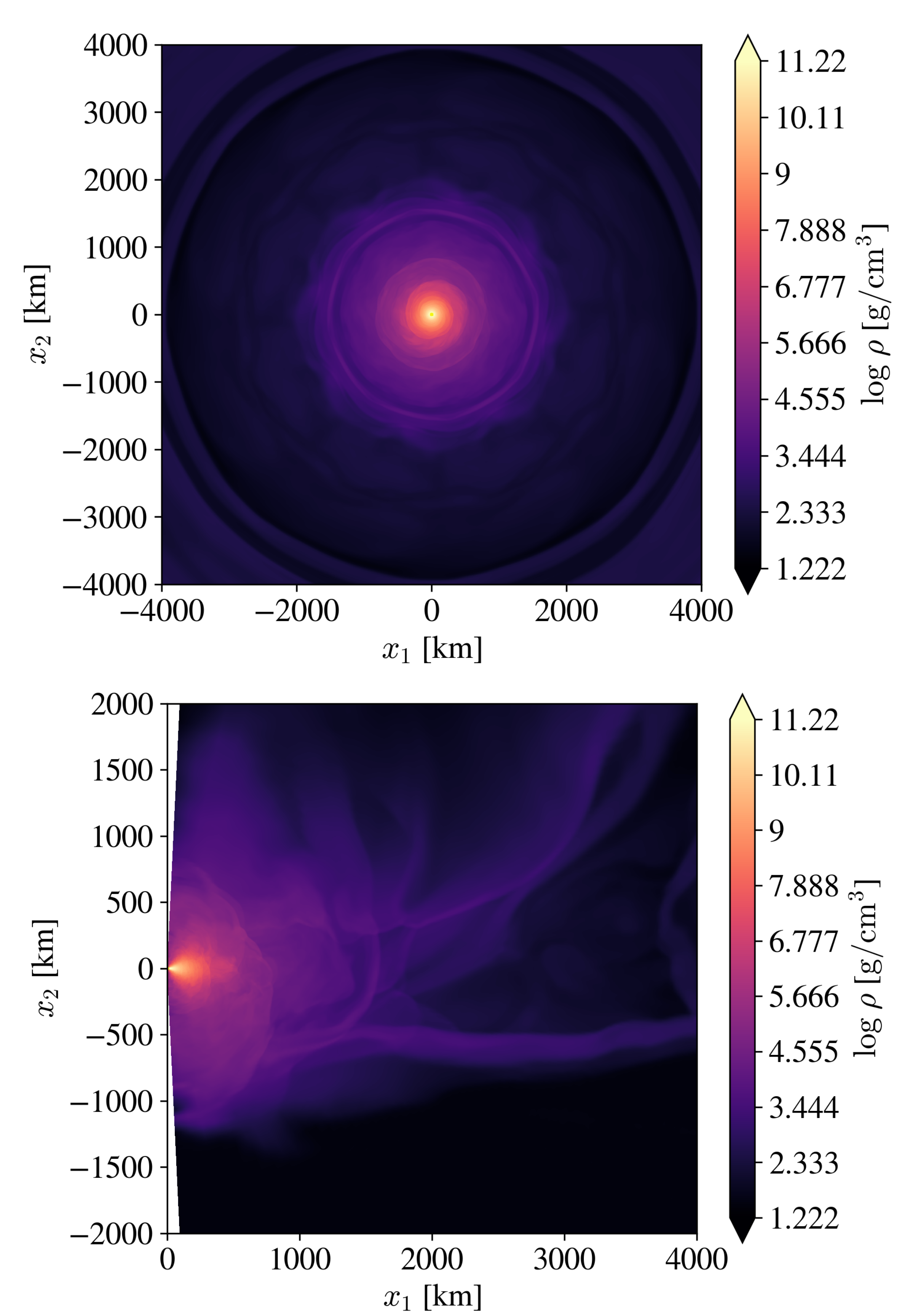}
\caption{Density of a magnetized torus including the impact of neutrinos at {114ms}. Shown is an equatorial cut (\textit{top} panel) and a meridional cut (\textit{bottom} panel). }
\label{fig:mhd_density}
\end{figure}

\begin{figure}[t!]
\includegraphics[scale=0.225]{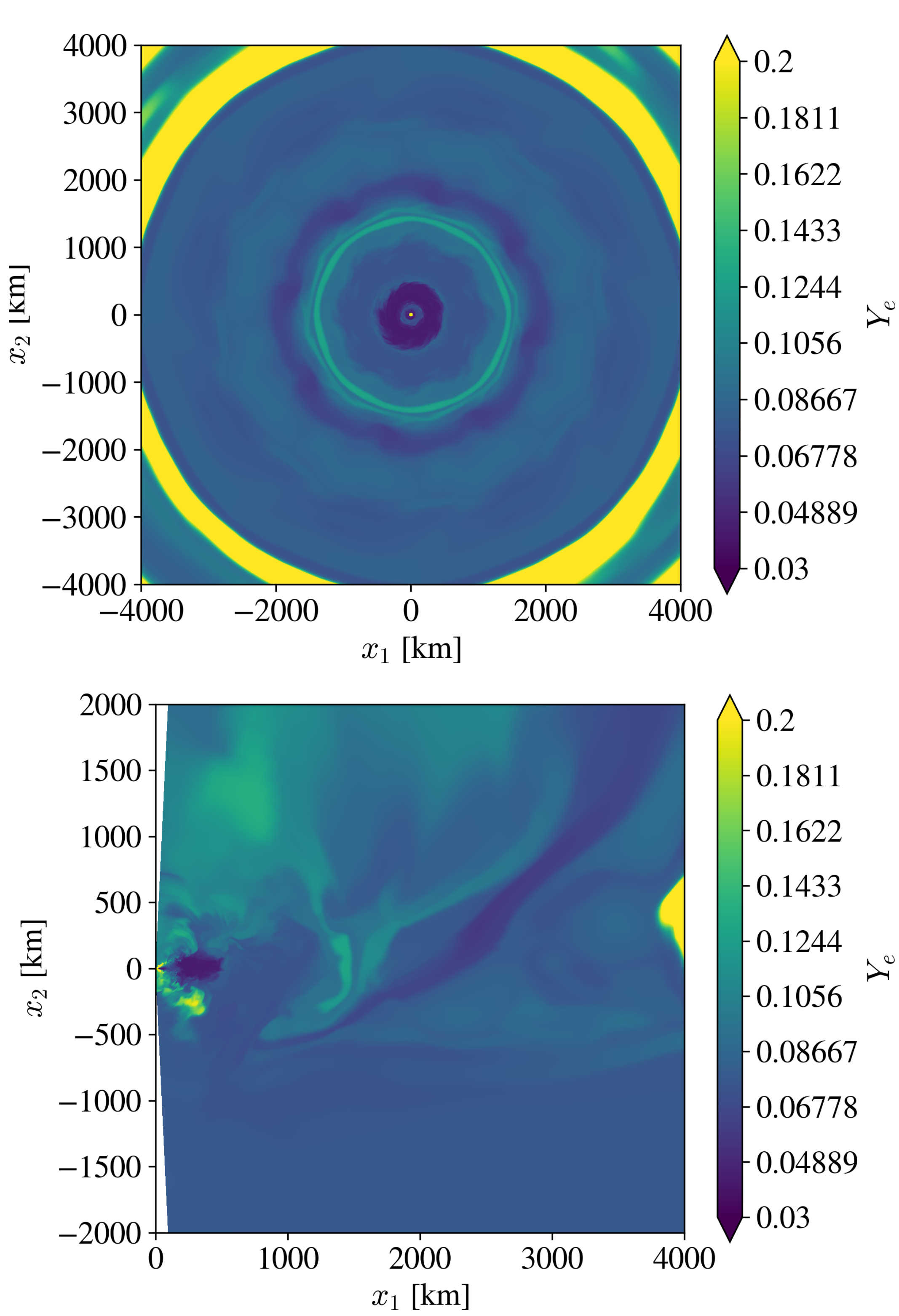}
\caption{Electron fraction of a magnetized torus including the impact of neutrinos  at {114ms}. Shown is an equatorial cut (\textit{top} panel) and a meridional cut (\textit{bottom} panel).}
\label{fig:mhd_ye_zoom}
\end{figure}

\begin{figure}[t!]
\includegraphics[scale=0.23]{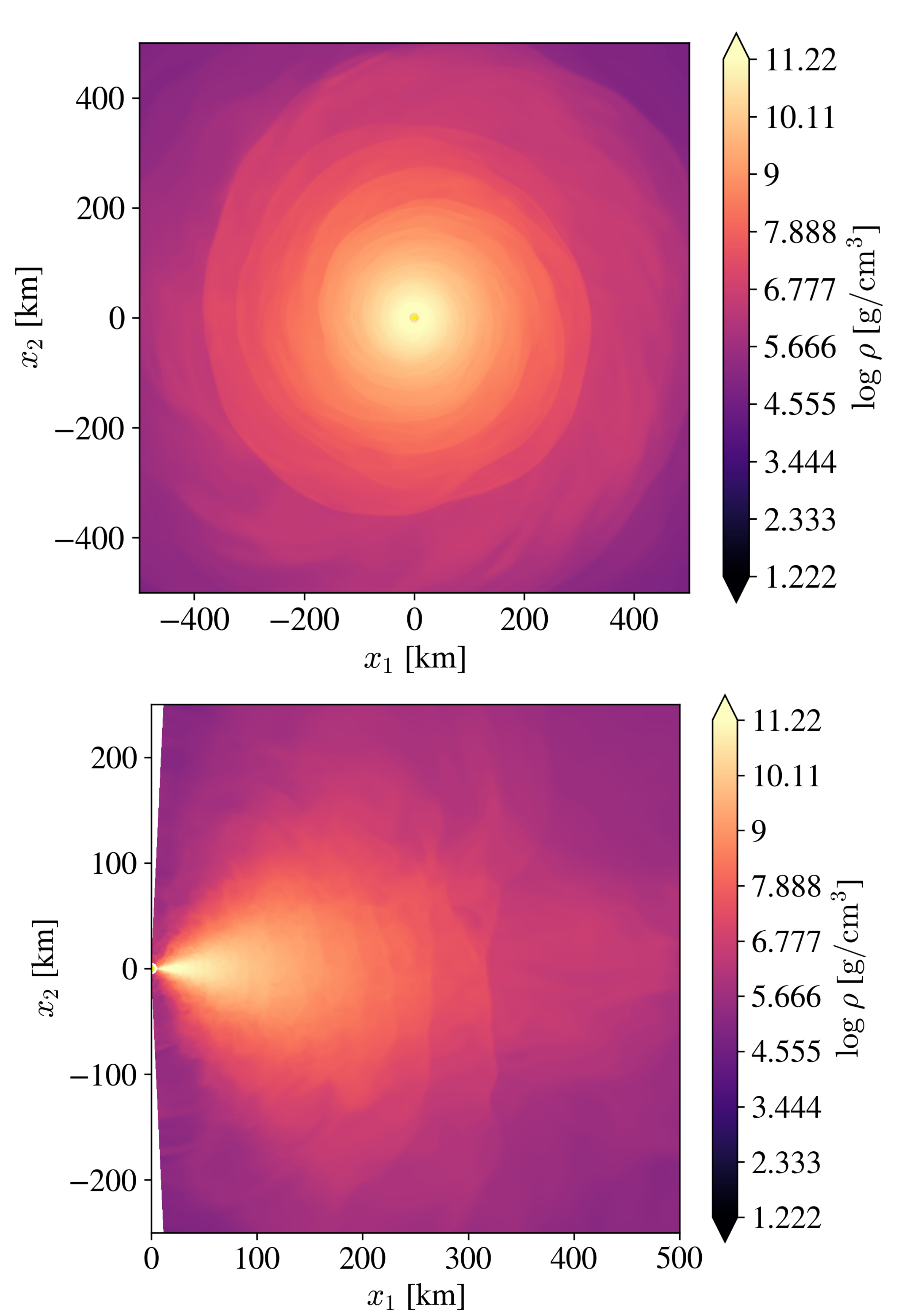}
\caption{Zoomed in version of the density of a magnetized torus including the impact of neutrinos  at {114ms}. Shown is an equatorial cut (\textit{top} panel) and a meridional cut (\textit{bottom} panel). }
\label{fig:mhd_density_zoom}
\end{figure}

\begin{figure}[t!]
\includegraphics[scale=0.225]{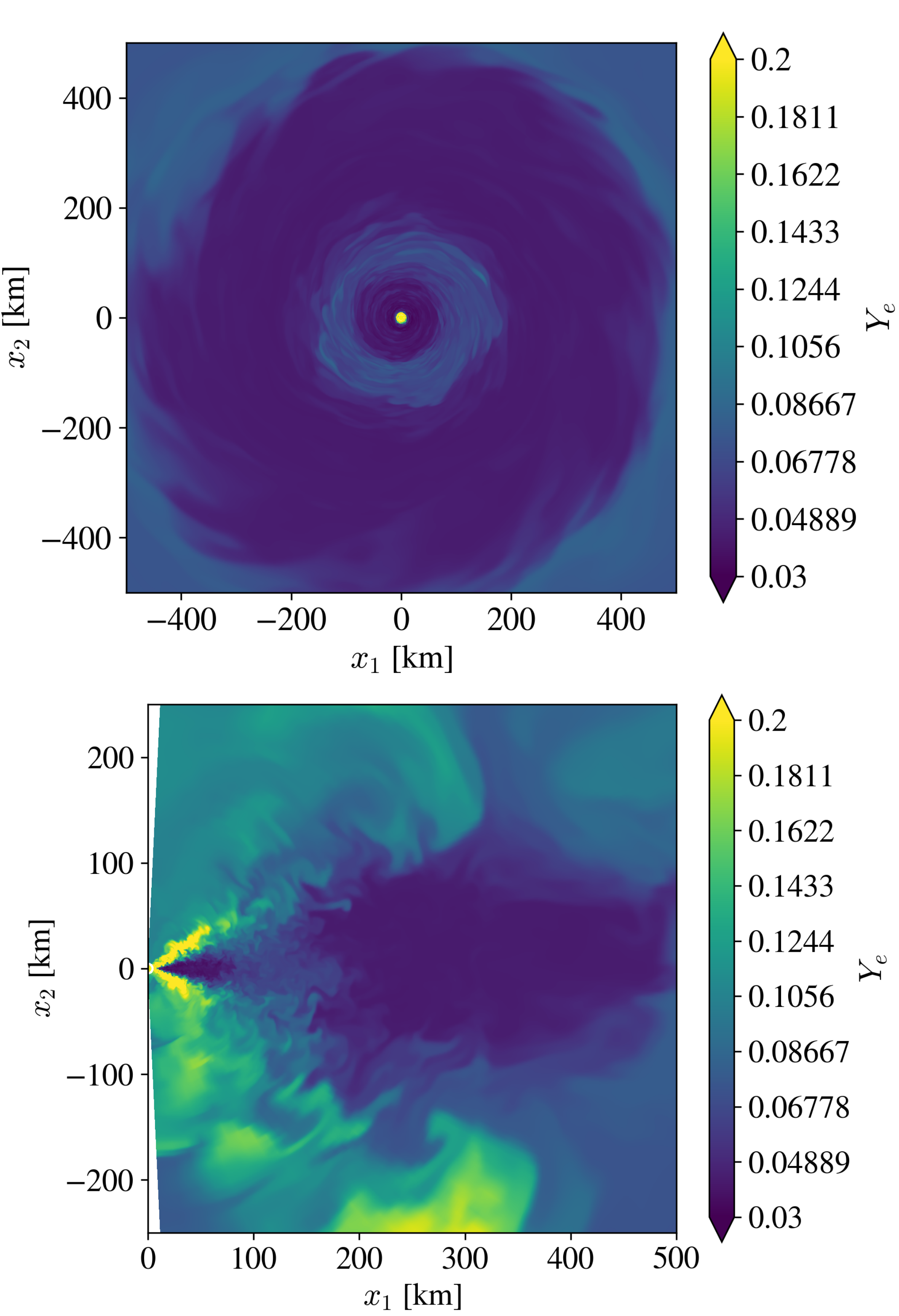}
\caption{Zoomed in version of the electron fraction of a magnetized torus including the impact of neutrinos  at {114ms}. Shown is an equatorial cut (\textit{top} panel) and a meridional cut (\textit{bottom} panel).}
\label{fig:mhd_ye}
\end{figure}

\section{Magnetized disk}
\label{mhd_torus}
In this section, we apply our new code HARM3D+NUC to a magnetized torus in 3d that approximates a post-merger disk. We use both the tabulated EOS and the leakage scheme in this test.

\subsection{Initial conditions}
The initial conditions inside the torus follow a similar setup to that of section~\ref{initial_hydro}, but with the addition of a poloidal magnetic field. {In order to start with a 
magnetic field devoid of
magnetic monopoles, we first set the vector potential to a prescribed distribution 
and calculate its curl using a finite difference operator compatible with our constrained transport method  \citep[see][for further details]{2014CQGra..31f5013Z}. Our poloidal magnetic field distribution results from a vector potential with only one non-zero component:
\beq{
A_\phi=\max\left( \overline{\rho}/\rho_{\rm max}-\rho_{0, \rm mag} \ , 0 \right)
} 
where $\overline{\rho}$ is the average density at that position, and $\rho_{\rm max}=1.66\times 10^{11} {\rm g/cm^3}$ is the maximum density of the torus. We set $\rho_{0, \rm mag}=0.2$ {in code units}, which corresponds to $\rho_{0, \rm mag}=3.33\times 10^{10}{\rm g/cm^3}$. 
Then we build the magnetic field with the vector potential and normalize its magnitude such that the ratio of the integrated gas pressure to integrated magnetic pressure is 100. Inside the disk, the matter is set to be neutron rich, $\Ye=0.1$.  The treatment of the atmosphere is the same as in section~\ref{atmosphere_torus}, except the density scales as $r^{-3/2}$. In the atmosphere, the electron fraction is set to its value in $\beta$-equilibrium, where the chemical potential of the neutrinos is set to zero. We show the parameters used in Table~\ref{table_mhd_torus}.

The simulations were performed in 3d on a grid designed to focus more cells 
about the equator and towards the black hole horizon.  We use the same grid as defined in \cite{2010ApJ...711..959N} but with different parameters.  The azimuthal grid spacing is uniform. 
The logarithmic radial grid is such that $\Delta r / r$ is fixed and the $i^\mathrm{th}$ cell center is located at: 
\beq{
r_i = \rmin \exp{\left[ \left( i + 1/2 \right) \log_\mathrm{10}{(\rmax/\rmin)} / N_r \right ] }
\ , 
}
with $\rmin = 1.303 r_{\rm g}$, $\rmax = 2000 r_{\rm g}$, 
and $i \in \left[0, N_r - 1\right]$.  The $\theta$ grid uses a high-order polynomial function to provide a nearly uniform grid spacing spacing near the equator: 
\beq{
\theta_j = \frac{\pi}{2} \left[ 1 + \left(1 - \xi \right) \left(2 x^{(2)}_j - 1 \right) +  \left( \xi - \frac{2 \theta_c}{\pi} \right) \left(2 x^{(2)}_j - 1 \right)^n 
\right] \ , 
}
where $\xi$ is a parameter controlling the severity of the focusing, $n$ is the order of polynomial used in the transformation, $\theta_c$ is the opening angle of the polar regions we excise,  $x^{(2)}_j \equiv \left(j + 1/2 \right) / N_\theta$, and $j \in \left[0, N_\theta - 1\right]$ .  In our run, 
we used $\theta_c = \pi 10^{-14}$, $\xi = 0.65$, and $n=7$. 
The number of cells per dimension used was $N_r \times N_\theta \times N_\phi = 1024\times 160 \times 256$. 

\begin{table}
\begin{tabular}{||c| c ||} 
 \hline
  Parameter   & Value  \\ [0.5ex] 
 \hline\hline
Disk radius of maximum pressure & $9 r_{\rm g}$ \\ 
Disk inner radius & $4 r_{\rm g}$ \\
Mass of disk & $0.03 M_{\odot}$  \\
$\Ye$ in the disk & 0.1  \\
Specific entropy in the disk & $7\ k_{\rm b}$/baryon  \\
{$(P_{\rm gas}/P_{\rm mag})$} & 100  \\
BH spin & 0.9375 \\
BH mass & $3 M_{\odot}$   \\
Specific enthalpy at boundary & 0.9977 [code units]  \\
{Temperature at radius of maximum pressure } & {4.4 MeV} \\
 [1ex] 
 \hline
 \end{tabular}
 \caption{Parameters used in the simulation.}
  \label{table_mhd_torus}
\end{table}

\subsection{Scaling tests}
We performed scaling tests for this run for 3 different number of processors: 5120, 2560 and 1280 processors. For this setup, the number of time steps in the code per second per processor were: 0.000723, 0.000781, 0.000868 respectively. The difference between 5120 and 1280 processors is around $17\%$.
If we do not include the neutrino leakage scheme but include only a tabulated EOS, for 2560 processors, the number of steps per second per processor is 0.001328, which makes the leakage $58\%$ slower than only considering the tabulated EOS. 

\subsection{Magnetic turbulence}

In order to confirm that we are adequately resolving magnetic turbulence, we display in Figure~\ref{fig:mhd_q_mri_colormap} the number of grid cells per wavelength of the fastest growing mode of the magneto-rotational instability (MRI), defined as 
\citep{2010ApJ...711..959N,2011ApJ...738...84H,2012ApJ...749..189S,2013ApJ...772..102H}:
\beq{
Q_{\rm mri, x}=\frac{\lambda_{x,\rm mri}}{\Delta_x}
}
where $x=\theta, \phi$, $\Delta_x$ is the cell size, and the wavelength of the fastest MRI growing mode is:
\beq{
\lambda_{x,\rm mri}=\frac{2\pi}{\Omega}\frac{|b^x|}{\sqrt{\rho h+b^2}} \ .
}

{ As can be seen in Fig~\ref{fig:mhd_q_mri}, our grid satisfies the criterion of \citet{2004ApJ...605..321S} everywhere  except for later times within $r<50$km.  While our results fail to meet criteria for asymptotic MRI convergence set forth in  \citet{2011ApJ...738...84H}, our disk does satisfy  $Q_{\rm mri, 3}>10$ everywhere, and $Q_{\rm mri, 2}>6$ for $r\gtrsim 50$km for most of the run. One reason why our simulation may not reach larger $Q_{\rm mri}$ values is because we used the same random perturbations across all MPI processes in the initial conditions. Because we used 16 subdomains in the azimuthal dimension, this means that the simulation is nearly periodic over $\Delta \phi = \pi/8$, and the  azimuthal modes with $m<8$ start off significantly weaker as they are seeded with perturbations at only the round-off error level. }

\begin{figure}[t!]
\includegraphics[scale=0.6]{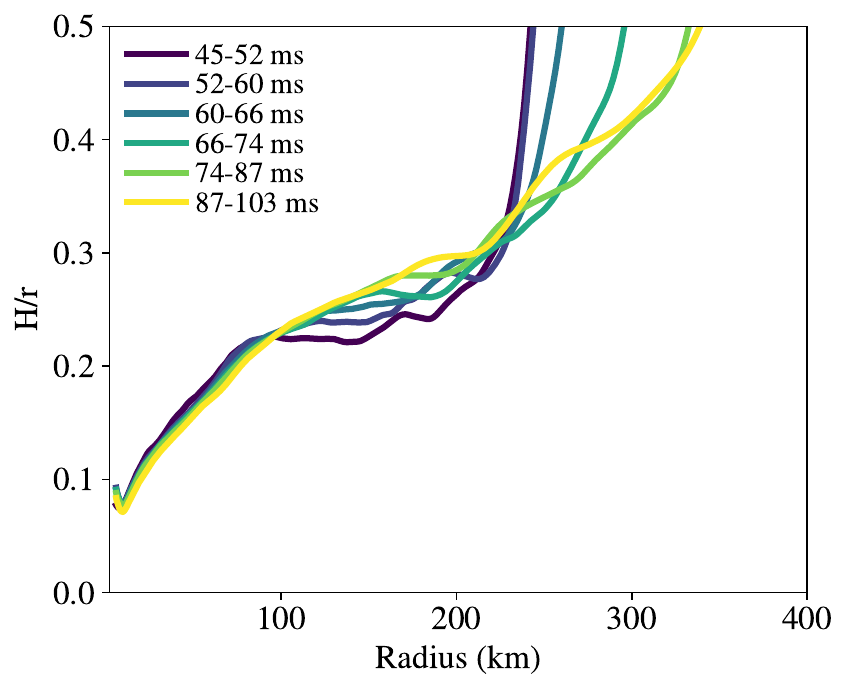}
\caption{Shown is the geometrical thickness ($H/r$) of the disk, as a function of radius. The thickness is averaged between the indicated time in the legend.}
\label{fig:mhd_h_o_r}
\end{figure}

\begin{figure}[t!]
\includegraphics[scale=0.57]{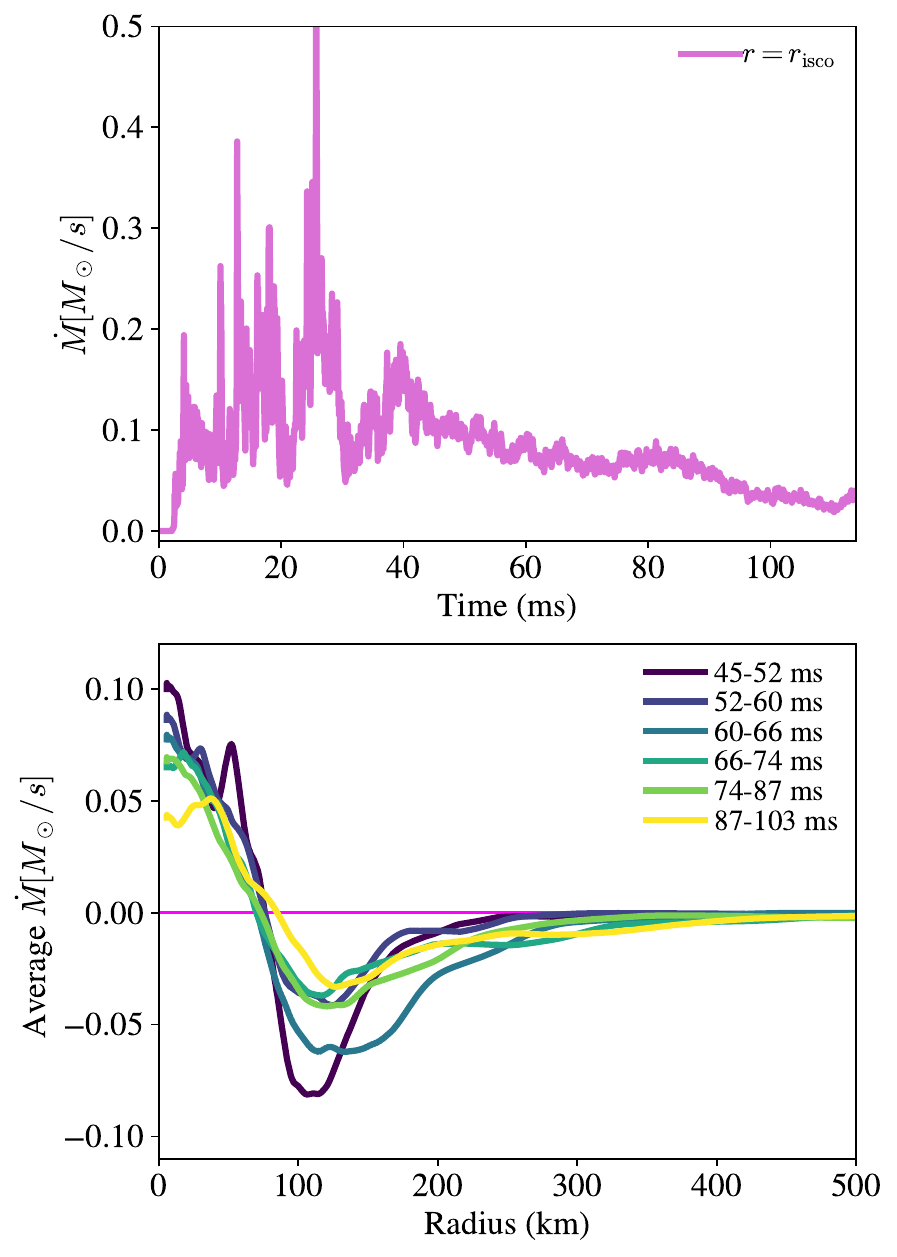}
\caption{\textit{Top} panel: Mass accretion rate onto the innermost stable circular orbit (ISCO) of the BH as a function of time. \textit{Bottom} panel: Average mass accretion rate as a function of radius. We averaged the mass accretion rate between the times indicated in the legend.}
\label{fig:mhd_mdot}
\end{figure}

\begin{figure}[t!]
    \includegraphics[scale=0.58]{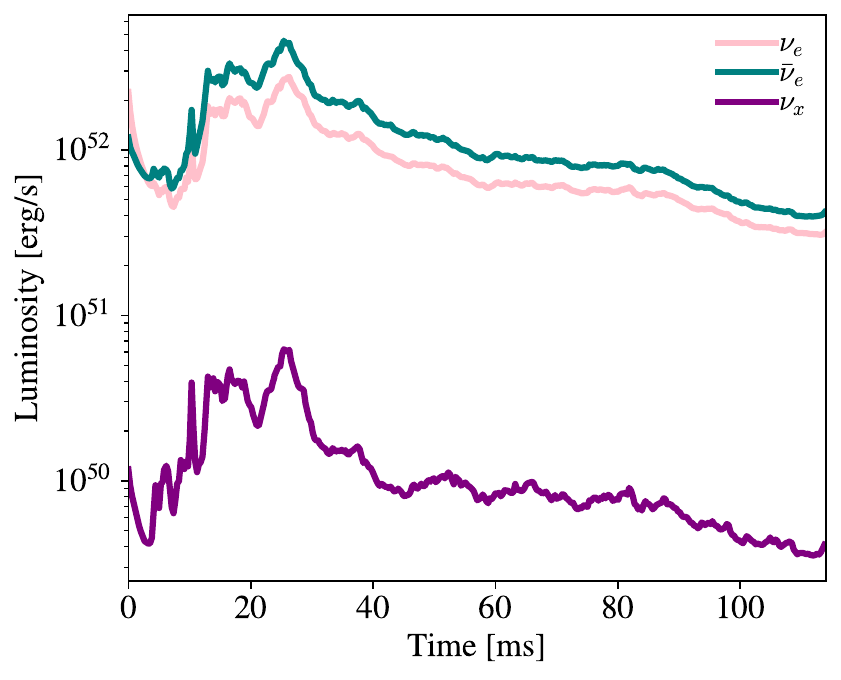}
\caption{Luminosity due to the different neutrino species as a function of time. }
\label{fig:lum_neutrinos}
\end{figure}

\subsection{Impact of neutrinos and EOS}
Magnetic stresses will transport angular momentum in the disk, heating the gas, which will produce a high velocity outflow \citep{2013MNRAS.435..502F,2018ApJ...858...52S}. This outflow will be affected by the addition of neutrinos formed through weak reactions. In the midplane, neutrinos will  carry significant amounts of energy, which will cool and make the disk {geometrically} thinner. Another outflow is also expected to occur in the outer regions of the disk due to the release in nuclear binding energy when there is recombination of free nucleons into $\alpha-$particles, which produces enthalpy and unbinds material \citep{2009ApJ...699L..93L,2013MNRAS.435..502F}. In this subsection we show the impact of both the emission of neutrinos and the recombination of free nucleons.

In Figures~\ref{fig:mhd_density} and~\ref{fig:mhd_ye} we display the outflows that results from our simulations of a neutrino-cooled magnetized disk at {114} ms. 
In Figures~\ref{fig:mhd_ye_zoom} and~\ref{fig:mhd_density_zoom} we plot the electron fraction and the density, respectively, at $t={114} \mathrm{ms}$. {Movies can be found \href{https://www.youtube.com/playlist?list=PLurnnzvqZvZaqLWlT2BVmPOlDm5P3BUOz}{here} \footnote{\url{https://www.youtube.com/playlist?list=PLurnnzvqZvZaqLWlT2BVmPOlDm5P3BUOz}}. }

{Neutrino cooling is expected to happen on the diffusion timescale, which is on the order of milliseconds, much shorter than our evolution timescale. } The inner regions of the disk are very neutron rich, confirming the self-regulating phase found in \citet{2017PhRvL.119w1102S,2018ApJ...858...52S}. In this phase, there is a balance between the neutrino cooling and the heating driven by MHD that self-regulates the electron degeneracy parameter, and the final state is a neutron rich disk \citep{2017PhRvL.119w1102S,2018ApJ...858...52S}. {We note that although this new code does not include neutrino absorption in the ejecta, absorption will modify the electron fraction in the outflow \citep{2021arXiv210208387J}. }

In the \textit{top} panel of Figure~\ref{fig:mhd_mdot} we show the mass accretion rate through the innermost stable circular orbit (ISCO)  as a function of time, and show the accretion rate as a function of radius in the \textit{bottom} panel. The outflow can be clearly seen as a negative mass accretion rate at larger radii, as well as a settling of the mass accretion rate as time passes.  

In Figure~\ref{fig:mhd_h_o_r},  we plot the geometrical thickness of the disk, or $H/r$. We estimated this thickness using the scale height $H$ following \citet{2012ApJ...755...51N}:
\beq{
H=\frac{\langle \rho\sqrt{g_{\theta \theta}}\left|\theta-\pi/2\right| \rangle}{\langle\rho\rangle}
}
where $\langle X \rangle$ is the average of the quantity $X$ over a spherical shell:
\beq{
\langle X \rangle = \frac{\int X \sqrt{-g} d\theta d\phi}{\int \sqrt{-g} d\theta d\phi }
\ . }
In the deepest regions of the disk, the heating due to MHD turbulence helps create neutrinos/anti-neutrinos, which escape, remove energy, and {geometrically} thin the disk.  Recombination of free nucleons into $\alpha$-particles releases binding energy,  
{effectively increasing the enthalpy and unbinds material.}
The effect of recombination is less severe than the {geometrically} thinning due to {neutrino/anti-neutrino losses}. This transition can be seen at around $150 \mathrm{km}$.

We may obtain the amount of energy radiated by each species of neutrino and anti-neutrino as was done in \citet{2018ApJ...858...52S}:
\beq{
L_{\nu_i}=\int \alpha\gamma \mathcal{Q}_{\nu_i}^{\rm eff} \sqrt{-g}d^3x
\ . }
In Figure~\ref{fig:lum_neutrinos}, we show the luminosity for each species. It can be seen that the electron neutrino (and anti-neutrino) dominate the emission over all of the other species of neutrino. The luminosity roughly follows the mass accretion rate as seen in Figure~\ref{fig:mhd_mdot}, as heating from the magnetic stresses ignite the creation of neutrinos/anti-neutrinos. This suggests the radiative efficiency of neutrino/anti-neutrinos emission remains relatively steady. 

Our initial conditions are similar (although not identical) to the initial conditions in \citet{2018ApJ...858...52S}. They performed 3d simulations of a post-merger accretion disk with a relatively higher specific entropy and lower spin than this simulation. They used Cartesian coordinates, a Helmholtz EOS for relatively low densities, and a neutrino leakage scheme. They evolved the disk for longer times (380 ms). Even though we use a different EOS (Sly4), the disk thickness is qualitatively similar. At the inner regions of the disk, neutrino cooling dominates, whereas at outer regions (at radius higher than around 100km), recombination  is responsible for {making the disk geometrically thicker}. The neutrino/anti-neutrino luminosities are comparable, \citet{2018ApJ...858...52S} has a higher luminosity, but that could be attributed to the difference in the initial disk specific entropy. 

As the outflow expands, it will cool, and heavy elements will be created via the $r-$process. We will explore this nucleosynthesis in a future paper. 

\section{Summary}
GRMHD simulations of post-merger accretion disks have advanced over the last few years with better treatment of neutrinos and a more realistic EOS. In this paper we present the addition of a neutrino leakage scheme and a tabulated EOS into the computationally efficient, versatile GRMHD code \harm. This new addition to \harm, called HARM3D+NUC, has the potential to be used in a range of simulations where neutrinos are present. In the paper, we use the new code HARM3D+NUC to simulate an accretion disk resembling the post-merger phase of a binary neutron star, though other applications include collapsars \citep[e.g., ][]  {2019Natur.569..241S,2020ApJ...902...66M}.

The paper shows how we implemented the tabulated EOS in the conserved variable to primitive variable routines, and the different methods we implemented and tested for performing this inversion. We show that using the 3d  primary recovery method is the most accurate and efficient, but least robust, choice which is why we also employ several 2d and 1d backup routines. The leakage scheme is implemented by adding the neutrino/anti-neutrino heating/cooling and emission/absorption terms as source terms in the equations of motion. We describe in detail an approach to obtain the optical depth locally and how we can use a convergence criterion to get the optical depth after a few iterations once the initial guess is made.

We show several tests for our new code. The tabulated EOS is tested by determining the relative error between original primitive variables and the recovered primitive variables. We also test the EOS by performing a simulation of a torus in hydrostatic equilibrium, showing that it stays in hydrostatic equilibrium throughout the entire simulation. We test the neutrino leakage scheme in the optically thin regime by investigating the $\beta-$process in a constant density gas. We  test the optical depth algorithm in a constant density circular disk and a stripes setup.

With our new machinery, we simulate a magnetized high-density torus, which serves as an approximation to the accretion flow after the merger of two neutron stars. Magnetic stresses transport angular momentum from the disk, driving a high velocity outflow. The outflow is affected by both the addition of neutrinos and the nuclear binding energy released from the recombination of nucleons to $\alpha-$particles, which acts to {geometrically thicken} the disk. Neutrinos will alter the electron fraction of the ejecta especially in the inner regions of the disk, whereas the recombination of nucleons is more prominent in the outer regions of the disk.  This highlights the importance of modeling  the accretion disk including neutrinos and an EOS that considers this extra unbinding of material due to recombination.

We plan to use the new code to do long-term evolutions of binary neutron star mergers starting from before the neutron stars merge to the evolution of the outflow. Heavy elements should be created via the $r-$process in this outflow as it expands and cools.  We plan to use different codes and methods to treat the initial data, pre-merger/merger, and post-merger phases. The initial data for the neutron stars will be constructed using a modified version of LORENE \citep{2016ascl.soft08018G} we have developed. Binaries will be evolved until they merge and eventually form a black hole surrounded by an accretion disk using two GRMHD codes: IllinoisGRMHD \citep{2015CQGra..32q5009E}, and Spritz \citep{2020CQGra..37m5010C}. {After the remnant has collapsed to a BH and the numerical metric has stabilized, we will interpolate the MHD primitives and numerical metric into the grid of HARM3D+NUC (L\'opez Armengol et al. in prep). After doing the appropriate tensorial transformations from the Cartesian base to the coordinate base of HARM3D+NUC, we will continue the post-merger evolution with HARM3D+NUC.} 


\begin{acknowledgements}

We thank the anonymous referee, R. Foley, B. Villase\~nor, D. Radice, T. Piran, V. Mewes, D. Siegel, S. Rosswog, A. Janiuk, J. Miller, J. Dolence, M.~C. Miller, P. M{\"o}sta, N.~M. Lloyd-Ronning, A. Batta, G. Koenigsberger, D. Kasen for valuable conversations. A.M-B and E. R-R are supported by the  Heising-Simons Foundation, the Danish National Research Foundation (DNRF132), NSF (AST-1911206 and AST-1852393). A.M-B is supported by the UCMEXUS-CONACYT Doctoral Fellowship. 
B.J.K was supported by the NASA Goddard Center for Research and Exploration in Space Science and Technology (CRESST) II Cooperative Agreement under award number 80GSFC17M0002. R.OS was supported by NSF PHY-2012057 and AST-1909534. This work was made possible by the NASA TCAN award TCAN-80NSSC18K1488. Computational resources were provided by the NCSA's Blue Waters sustained-petascale computing NSF projects OAC-1811228 and OAC-1516125,  by the TACC's Frontera NSF projects  PHY20010 and AST20021, and the lux supercomputer at UC Santa Cruz, funded by NSF MRI grant AST 1828315.

\end{acknowledgements}

\bibliography{neutrinos.bib}

\end{document}